%% file: Zp.tex
\documentclass[11pt,a4paper]{article}

\usepackage{amsmath}

\usepackage{cite}
\usepackage{jheppub}

\RequirePackage{graphicx}
\usepackage{atlasphysics}
\usepackage{lldefs}
\usepackage{preprintcover}  

\input{Numbers}

\PreprintCoverPaperTitle{Search for high-mass resonances decaying to dilepton final states in $pp$ collisions at $\rts = 7$~TeV with the ATLAS detector}  
\PreprintIdNumber{CERN-PH-EP-2012-190}  
\PreprintCoverAbstract{The ATLAS detector at the Large Hadron Collider is used to search for high-mass resonances
decaying to an electron-positron pair or a muon-antimuon pair. 
The search is sensitive to 
heavy neutral \zp\ gauge bosons, Randall-Sundrum gravitons, \zstar\ bosons, techni-mesons, Kaluza-Klein $Z/\gamma$ bosons, 
and  bosons predicted by Torsion models.  
Results are presented based on an analysis of $pp$ collisions at a center-of-mass energy of 7~TeV 
corresponding to an integrated luminosity of \LumiElectronsFb~\ifb\ in the \ee\  channel
and \LumiMuonsFb~\ifb\ in the \mumu channel.
A \zp\ boson with Standard Model-like  couplings is excluded at 95\% confidence level for masses below 
 \LimitCombined~TeV. 
A Randall-Sundrum graviton with coupling $\kovermb=0.1$ is excluded at 95\% confidence level for masses below 
\LimitCombinedG~TeV. 
Limits on the other models are also presented, including Technicolor and \MM.
}  
\PreprintJournalName{JHEP}  

\begin{document}

\title{Search for high-mass resonances decaying to dilepton final states in $pp$ collisions at $\rts = 7$~TeV with the ATLAS detector}

\author{The ATLAS Collaboration}

\abstract{
The ATLAS detector at the Large Hadron Collider is used to search for high-mass resonances
decaying to an electron-positron pair or a muon-antimuon pair. 
The search is sensitive to 
heavy neutral \zp\ gauge bosons, Randall-Sundrum gravitons, \zstar\ bosons, techni-mesons, Kaluza-Klein $Z/\gamma$ bosons, 
and  bosons predicted by Torsion models.  
Results are presented based on an analysis of $pp$ collisions at a center-of-mass energy of 7~TeV 
corresponding to an integrated luminosity of 
\LumiElectronsFb~\ifb\ in the \ee\  channel
and \LumiMuonsFb~\ifb\ in the \mumu channel.
A \zp\ boson with Standard Model-like  couplings is excluded at 95\% confidence level for masses below 
 \LimitCombined~TeV. 
A Randall-Sundrum graviton with coupling $\kovermb=0.1$ is excluded at 95\% confidence level for masses below 
\LimitCombinedG~TeV. 
Limits on the other models are also presented, including Technicolor 
and \MM.
}

\maketitle

\newpage

\section{Introduction}
\input{intro}

\section{ATLAS detector}
\input{detector}

\section{Lepton reconstruction}
\input{e_reco}

\input{mu_reco}

\section{Event selection}
\input{e_selection}
\input{mu_selection}
\input{exp_bg}

\input{simulation}

\input{data_driven}

\section{Systematic uncertainties}
\input{systematics}

\input{results}
\section{Limit-setting procedure}
\input{simple_limits}

\input{limits_with_interf}
\section{Conclusions}
\input{conclusion}

\section{Acknowledgements}
\input{acknowl}

\bibliographystyle{JHEP}

\providecommand{\href}[2]{#2}\begingroup\raggedright\endgroup

\newpage
\input{atlas_authlist}

\end{document}

%% file: Numbers.tex
\newcommand{\LumiElectronsFb}{4.9}
\newcommand{\LumiMuonsFb}{5.0}

\newcommand{\pvalele}{36\%}
\newcommand{\pvalmuon}{68\%}
\newcommand{\pvalcomb}{40\%}

\newcommand{\LimitElectronExpected}{2.13}
\newcommand{\LimitElectron}{2.08}

\newcommand{\LimitMuonExpected}{2.00}
\newcommand{\LimitMuon}{1.99}

\newcommand{\LimitCombinedExpected}{2.25}
\newcommand{\LimitCombined}{2.22}

\newcommand{\LimitCombinedExpectedSq}{1.95}
\newcommand{\LimitCombinedSq}{1.91}

\newcommand{\LimitCombinedExpectedN}{1.87}
\newcommand{\LimitCombinedN}{1.79}

\newcommand{\LimitCombinedExpectedPsi}{1.87}
\newcommand{\LimitCombinedPsi}{1.79}

\newcommand{\LimitCombinedExpectedChi}{2.00}
\newcommand{\LimitCombinedChi}{1.97}

\newcommand{\LimitCombinedExpectedEta}{1.92}
\newcommand{\LimitCombinedEta}{1.87}

\newcommand{\LimitCombinedExpectedI}{1.91}
\newcommand{\LimitCombinedI}{1.86}

\newcommand{\LimitElectronExpectedZs}{2.13}
\newcommand{\LimitElectronZs}{2.10}

\newcommand{\LimitMuonExpectedZs}{1.99}
\newcommand{\LimitMuonZs}{1.97}

\newcommand{\LimitCombinedExpectedZs}{2.22}
\newcommand{\LimitCombinedZs}{2.20}

\newcommand{\LimitElectronExpectedG}{2.04}
\newcommand{\LimitElectronG}{2.03}

\newcommand{\LimitMuonExpectedG}{1.93}
\newcommand{\LimitMuonG}{1.92}

\newcommand{\LimitCombinedExpectedG}{2.17}
\newcommand{\LimitCombinedG}{2.16}
\newcommand{\limitGraviton}{\LimitCombinedG}
\newcommand{\limitGravitonExp}{\LimitCombinedExpectedG}

\newcommand{\LimitCombinedGOne}{0.92}
\newcommand{\LimitCombinedGThree}{1.49}
\newcommand{\LimitCombinedGFive}{1.72}
\newcommand{\LimCombGTwelve}{2.23}
\newcommand{\LimCombGFourteen}{2.32}
\newcommand{\LimCombGSeventeen}{2.42}
\newcommand{\LimCombGTwenty}{2.51}

\newcommand{\LimitCombinedGExpOne}{1.02}
\newcommand{\LimitCombinedGExpThree}{1.53}
\newcommand{\LimitCombinedGExpFive}{1.81}
\newcommand{\LimCombGExpTwelve}{2.25}
\newcommand{\LimCombGExpFourteen}{2.33}
\newcommand{\LimCombGExpSeventeen}{2.44}
\newcommand{\LimCombGExpTwenty}{2.53}

\newcommand{\LimCombExpTSHalf}{1.58}
\newcommand{\LimCombTSHalf}{1.52}

\newcommand{\LimCombExpTSOne}{1.96}
\newcommand{\LimCombTSOne}{1.94}

\newcommand{\LimEleExpTSTwo}{2.20}
\newcommand{\LimEleTSTwo}{2.15}

\newcommand{\LimMuExpTSTwo}{2.08}
\newcommand{\LimMuTSTwo}{2.07}

\newcommand{\LimCombExpTSTwo}{2.31}
\newcommand{\LimCombTSTwo}{2.29}

\newcommand{\LimCombExpTSThree}{2.55}
\newcommand{\LimCombTSThree}{2.50}

\newcommand{\LimCombExpTSFour}{2.77}
\newcommand{\LimCombTSFour}{2.69}

\newcommand{\LimCombExpTSFive}{3.02}
\newcommand{\LimCombTSFive}{2.91}

\newcommand{\LimitElectronExpLSTC}{0.85}
\newcommand{\LimitElectronLSTC}{0.85}

\newcommand{\LimitMuonExpLSTC}{0.71}
\newcommand{\LimitMuonLSTC}{0.70}

\newcommand{\LimitCombinedExpLSTC}{0.89}
\newcommand{\LimitCombinedLSTC}{0.85}

\newcommand{\LimCombMASix}{359}
\newcommand{\LimCombExpMASix}{352}
\newcommand{\LimCombMAFive}{485}
\newcommand{\LimCombExpMAFive}{516}
\newcommand{\LimCombMAFour}{768}
\newcommand{\LimCombExpMAFour}{742}
\newcommand{\LimCombMAThree}{1175}
\newcommand{\LimCombExpMAThree}{1233}
\newcommand{\LimCombMATwo}{1566}
\newcommand{\LimCombExpMATwo}{1605}

\newcommand{\LimitElectronExpKKgf}{3.11}
\newcommand{\LimitElectronKKgf}{3.35}

\newcommand{\LimitMuonExpKKgf}{3.38}
\newcommand{\LimitMuonKKgf}{3.55}

\newcommand{\LimitCombinedExpKKgf}{4.07}
\newcommand{\LimitCombinedKKgf}{4.16}

\newcommand{\LimitElectronExpKKgsq}{3.52}
\newcommand{\LimitElectronKKgsq}{4.03}

\newcommand{\LimitMuonExpKKgsq}{3.79}
\newcommand{\LimitMuonKKgsq}{3.93}

\newcommand{\LimitCombinedExpKKgsq}{4.53}
\newcommand{\LimitCombinedKKgsq}{4.71}



%% file: intro.tex
Searches for new resonances 
decaying to a dilepton final state 
have had a long and successful history. These channels contributed to the
 discovery of the  quarkonium resonances $J/\psi$ and
 $\Upsilon$, as well as the discovery of the $Z$ boson. Various models beyond the Standard Model (SM) contain additional bosons which can decay
 into dileptons, providing a fully reconstructable final state with small, well-understood backgrounds.
In this article, data collected by the ATLAS experiment at the LHC are used to search for 
new resonances decaying 
into  dielectron and dimuon final states. 

The Sequential Standard Model (SSM)~\cite{Langacker:2008yv} defines the \zpssm\ couplings to SM fermions to be the same as 
the SM $Z$ boson couplings and
is often used in the literature as a benchmark model. 
A number of models predict additional neutral vector gauge bosons. 
One class postulates larger symmetry groups in which the SM gauge group is embedded. 
This is usually motivated by gauge unification
 or restoration of left-right symmetry, which is violated by the weak interaction. In one scheme, the SM gauge group derives from the $E_6$ group 
 which, upon symmetry-breaking via the $SU(5)$ subgroup, results in two additional $U(1)$ gauge groups named $U(1)_\chi$ and $U(1)_\psi$
with associated gauge bosons \zpchi\ and \zppsi\ that can mix~\cite{London:1986dk,Langacker:2008yv}. 
In the \MM~\cite{Villadoro}, 
 the phenomenology is controlled by only two effective coupling constants in addition to the \zp\ boson mass.
This parameterization encompasses many models, including 
a left-right symmetric model~\cite{Senjanovic:1975rk,Mohapatra:1974hk} 
and the pure (B--L) model~\cite{Basso:2008iv}, 
where B~(L) is the baryon (lepton) number, and B--L is the conserved quantum number. 

 A second set of models is motivated by various solutions to the hierarchy problem of the SM relating the very different scales of electroweak symmetry breaking and the gravitational Planck scale (\mpl).
One  class of such models introduces a new doublet of vector bosons $(\zstar, \wstar)$~\cite{wzstar} 
with masses not far from the weak scale~\cite{wzstar_motivate}, which couple to SM fermions only via magnetic-type interactions.
Compared to \zp\ bosons, interactions mediated by \zstar\ bosons are additionally suppressed in low-energy processes 
by powers of the small momentum transfer. Thus, the search for the \zstar\ boson is well-motivated at the LHC.

An alternative solution to the hierarchy problem has been proposed in models that 
allow the gravitational force to propagate into extra spatial dimensions. 
Among them, the Randall-Sundrum (RS) model~\cite{RS} 
predicts a warped space-time metric in one extra dimension.
Due to  warping,  the apparent strength of gravity in the four-dimensional subspace populated by the SM particles is 
exponentially suppressed. 
The RS model predicts excited states 
 of the graviton, \gstar , whose couplings to the SM particles are not exponentially suppressed.
The graviton is a spin-2 boson 
that can decay into dilepton final states 
with a coupling strength of \kovermb, 
where $k$ is a scale that defines the warp factor of the extra dimension, 
and 
$\mplb =\mpl /\sqrt{8\pi}$.

In Kaluza-Klein TeV$^{-1}$ models~\cite{Antoniadis:1990ew,Antoniadis:1999bq,Bella:2010sc}, 
the extra-dimensional momentum is  quantized by the inverse of the size of the extra dimension, 
creating a tower of massive Kaluza-Klein (KK) states corresponding to each SM particle. 
The Kaluza-Klein towers corresponding to the photon and the $Z$ boson, 
 \gkk\ and $Z_{\rm KK}$, 
 would manifest themselves as 
nearly degenerate 
resonances decaying to dilepton final states. 
This work is the first direct search for these Kaluza-Klein states. Previous bounds on the Kaluza-Klein  boson mass
were obtained from indirect measurements~\cite{Rizzo:1999en,PDG}.

 Technicolor models~\cite{TC-Weinberg,TC-Susskind,TC1}   
provide a dynamical scenario of Standard Model electroweak symmetry breaking (EWSB) by postulating a new strong binding force between techni-fermions.
 This model predicts additional bound states, techni-mesons,
 which  are resonances with masses of a few hundred GeV that can decay into fermion-antifermion pairs. 
Two main Technicolor models have a well-developed phenomenology at LHC energies. In Low-scale Technicolor (LSTC)~\cite{TC2,TC6}, the coupling constant varies slowly (walks) due to the existence of many scales of strong interactions while the phenomenology is dictated by the lowest mass particles. Minimal Walking Technicolor (MWT)~\cite{TC3,TC4,TC5} is a minimal model that is conformal and satisfies electroweak  precision measurements.

One of the main limitations of the Standard Model  is its inability to incorporate gravity. 
To address this problem, a consistent quantum theory of all four fundamental forces should 
be developed to seamlessly unify the SM and General Relativity (GR). 
However, so far no generally accepted formulation of quantum gravity exists, and therefore 
it is common to apply a phenomenological approach to the problem by considering extensions of GR 
and assuming that they might arise  from a more fundamental theory, such as String Theory. 
Among  such  extensions  
is gravity with Torsion~\cite{Shapiro:2001rz}.
In Torsion models, the spin of the elementary particles is the source of an extra field called Torsion, 
which interacts with SM fermions~\cite{Belyaev:2007fn,deAlmeida:2008dk}.   
This article reports on the first interpretation of a high mass dilepton search in terms of a Torsion resonance. 
 
Previous searches have set direct and indirect constraints on the mass of new heavy resonances~\cite{resonanceReview,Langacker:2009su}.
The Tevatron data~\cite{Abazov:2010ti,CDF:Zpmumu} have excluded a \zpssm\ boson with a mass lower than 1.071~TeV~\cite{CDF:Zpmumu}. 
Recent measurements from the LHC experiments~\cite{Aad:2011eps,CMS_Zp2011}, 
based on up to 5~\ifb\ of data, have excluded a \zpssm\ boson with a mass lower than 2.33~TeV~\cite{CMS_Zp2011}. 
Indirect constraints from LEP~\cite{Abbiendi:2003dh,Abdallah:2005ph,Achard:2005nb,Schael:2006wu}
have excluded \zpssm\ bosons with mass less than 1.787~TeV~\cite{Langacker:2009su}.
Constraints on the mass of the RS graviton have been set by the ATLAS~\cite{ATLAS_grav_diphoton}, CMS~\cite{CMS_Zp2011},
CDF~\cite{Aaltonen:2011xp} and D0~\cite{Abazov:2010xh} collaborations, 
excluding RS gravitons with a mass less than 2.14~TeV for $\kovermb=0.1$~\cite{CMS_Zp2011}. 
A \zstar\ with mass less than 1.152~TeV has been excluded by ATLAS~\cite{Aad:2011xp}. 
A search for the techni-mesons \rhot\ and \omegat\ in the dilepton final state has been conducted by CDF, 
resulting in a lower bound on the  \rhot\ and \omegat\ masses of 280~GeV~\cite{techniCDF}. 
The constraints from electroweak precision measurements give a lower limit on the \zkk\ boson masses around 4~TeV~\cite{Rizzo:1999en,GG}.

The results reported in this article use the full data sample recorded by ATLAS in 2011, corresponding to 
a total integrated luminosity of \LumiElectronsFb\  (\LumiMuonsFb)~\ifb\ 
in the \ee\ (\mumu) channel.

%% file: detector.tex
The ATLAS detector~\cite{atlas:detector}  
consists of an inner tracking detector surrounded by a 2~T superconducting solenoid, 
electromagnetic and hadronic calorimeters, and a muon spectrometer.
Charged particle tracks
in the pseudorapidity\footnote{ATLAS uses a right-handed coordinate system with the $z$-axis along the beam pipe. 
The $x$-axis points to the centre of the LHC ring, and the $y$ axis points upward. 
Cylindrical coordinates $(r,\phi)$ are used in the transverse $(x,y)$ plane, $\phi$ being the azimuthal angle. 
The pseudorapidity is defined in terms of the polar angle $\theta$ as $\eta=-\ln\tan(\theta/2)$.
} 
range $|\eta| < 2.5$ are reconstructed with the inner detector, which consists of silicon pixel, silicon strip, and transition radiation detectors.
The superconducting solenoid is surrounded by a hermetic calorimeter that covers $|\eta| < 4.9$.
For $|\eta| < 2.5$, the electromagnetic calorimeter is finely segmented and plays an important role in electron identification.
Outside the calorimeter, air-core toroids provide the magnetic field for the muon spectrometer. 
Three stations of precision drift tubes (with cathode strip chambers for the innermost station for $|\eta|>2.0$)
provide an accurate measurement of the muon track curvature in the range $|\eta |< 2.7$.
Resistive-plate and thin-gap chambers provide muon triggering capability in the range $|\eta |< 2.4$.

%% file: e_reco.tex


Electron candidates are 
formed from clusters of cells reconstructed in the electromagnetic calorimeter 
that are associated with a charged particle track in the inner detector. 
Measurements of the transverse calorimeter shower shape 
and the longitudinal leakage into the hadronic compartment~\cite{atlas:egamma_perf} 
are also used to 
improve electron-hadron identification.


The energy of an electron is obtained from the calorimeter, and its direction from the associated track.
At large transverse energy (\et ), the calorimeter energy resolution is dominated  by a constant term
 which is measured in data to be 1.2\% in the barrel ($|\eta| < 1.37$) and 1.8\% in the endcaps ($1.52 < |\eta| \leq 2.47$)~\cite{atlas:egamma_perf}.
For dielectron masses above 200~GeV, the mass resolution is below 2\% over the entire $\eta$ range.

%% file: mu_reco.tex

Muon tracks are first reconstructed separately in the inner detector (ID) 
and in the muon spectrometer (MS). 
The two tracks are then  matched and a combined fit is performed to the inner detector and muon spectrometer hits, 
taking into account the effect of multiple scattering and energy loss in the calorimeters.

The muons used in this work have hits in either three or two (out of three) stations of the muon spectrometer.  
Muons with hits in three stations, referred to as tight muons, comprise about 95\% of the sample, and have transverse momentum (\pt ) 
resolution at 1~TeV ranging from 10\% to 25\%.  Muons with hits in two stations, referred to as loose muons, have slightly worse \pt\ 
resolution than the tight muons.  Loose muons are accepted only in the barrel region of the muon spectrometer ($|\eta| < 1.05$), excluding small geometrical 
regions where the detector alignment is known to be less precise.

%% file: e_selection.tex

The data used for this study are required to have been recorded during periods of stable LHC beams, and when all relevant systems of the detector were operating normally.
Collision candidates are selected by requiring a primary vertex with at least three associated charged particle tracks, each with \pt\ more than 0.4~\gev. 

In the \ee\ channel,
events were triggered by a diphoton trigger, requiring
the presence of two electromagnetic clusters fulfilling a set of requirements~\cite{ATLASphotons} on the shape of the energy deposit
and with a transverse energy threshold of 20~GeV.
The efficiency of the diphoton trigger was measured in data to be 99\% 
for electron pairs forming dilepton masses above 100~GeV.
This was done using a tag-and-probe method on electrons from decays of \z\ bosons, 
selected using a single-electron trigger and requiring two electrons in the event to pass the event selection described below.
Since the trigger signals are saturated for electromagnetic clusters with very high energies, the trigger-level bunch-crossing identification, which uses the pulse shape, is challenging and performed by a dedicated algorithm, implemented in the first-level calorimeter trigger hardware.

Dielectron events are selected by requiring two electron candidates with the \medium\ level of identification defined 
in ref.~\cite{atlas:egamma_perf}, with transverse energy \et\ larger than 25~\gev\ and $|\eta| \leq 2.47$; 
the transition region between the barrel and endcap calorimeters is excluded.
A hit in the first layer of the pixel detector is required if an active pixel module is traversed, to suppress background from photon conversions.
To suppress background from QCD multijet production,  
the electron with the higher \et\ must be isolated, 
requiring $\Sigma \et(\Delta R<0.2)$ less than 7~\gev, where $\Delta R=\sqrt{(\Delta \eta)^2 + (\Delta \phi)^2}$ 
and  $\Sigma \et (\Delta R < 0.2)$ is the sum of the transverse energies in calorimeter cells around the electron direction 
in a cone of $\Delta R$ smaller than 0.2. 
The sum excludes the core of the electron energy deposition and is corrected for transverse shower leakage and pile-up from additional $pp$ collisions. 
The two highest \et\ electrons passing the above selection criteria are used to reconstruct the dielectron candidate.
The curvature measured by the inner detector for the high-energy electrons relevant to this analysis is not large enough to 
allow a precise determination of the transverse momentum and charge of the electrons. 
To avoid losses in efficiency,  the two electron candidates are not required to have opposite charge.
For the selection criteria described above and dielectron invariant masses (\mepem) greater than 130~GeV, the overall event acceptance times efficiency (\acc) for a \zp\ boson of mass 2~\tev\ is about 66\%.

%% file: mu_selection.tex

In the \mumu\ channel, 
events were triggered by at least one of two single-muon triggers, 
one with a \pt\ threshold of 22~GeV as reconstructed from the combination of ID and MS information, 
and the second with a \pt\ threshold of 40~GeV as reconstructed by the MS in the barrel region only.
The typical single-muon trigger efficiency was measured in data to be 85\% in the barrel (considering the union of both trigger paths) 
and 86\% in the endcaps.
The trigger efficiency is lower for muons than it is for electrons because of the smaller geometrical acceptance of the muon trigger detectors.

A dimuon event candidate is constructed from two opposite-charge muons, each with \pt\ greater than 25~GeV, $|\eta|<2.4$. In order to reject muons from cosmic radiation, the impact parameter with respect to the primary vertex must be smaller than 0.2 mm in the transverse plane and 1.0 mm along the beam axis, and the primary vertex must be reconstructed within 20 cm from the centre of the detector along the beam direction.
To ensure good \pt\ resolution, each muon is required to have a minimum 
number of hits in each of the inner detector components
as well as in three (two) muon spectrometer stations for tight (loose) muons. 
Muon candidates are excluded from the analysis if they cross regions of the muon spectrometer in which the bending power of the magnetic field is rapidly changing with the track position or the detector is less precisely aligned or calibrated. 
In addition, the difference between the standalone momentum measurements  
from the inner detector  and the muon spectrometer must not exceed five (three) times 
the sum in quadrature of the standalone resolutions 
for tight (loose) muons.
Finally, to suppress background from QCD multijet production,
each muon must be isolated,
requiring the sum of the \pt\ of all other tracks 
in a cone of size $\Delta R < 0.3$ to be less than 5\% of the transverse momentum of the muon.

Dimuon event candidates with two tight muons are considered first (tight dimuon selection). 
If more than one
such pair  
is found in an event, 
the one with the highest scalar sum of the leptons' \pt\ is selected.
If no tight muon pair is found, pairs with one tight muon and one loose muon
(loose dimuon selection) are considered.  Similarly, if more than one loose muon pair 
is found in an event, the one with the highest $\sum{|\pt|}$ is selected.
For the selection criteria described above, the overall event \acc\ for a \zp\ boson of mass 2~\tev\ decaying into a dimuon final state is 43\%,
including 4\% from the loose dimuon selection. 
The lower acceptance compared to the dielectron channel is due to the stringent hit  requirements in the muon spectrometer. 

%% file: exp_bg.tex

For both channels, the dominant and irreducible background is due to 
the $Z/\gamma^*$ (Drell-Yan) process, characterized by the same final state as the signal.  
Small contributions from \ttbar\ and diboson ($WW$, $WZ$ and $ZZ$) production are also 
present in both channels. Semi-leptonic decays of $b$ and $c$ quarks in the \ee\ and \mumu\ samples, plus a mixture of photon conversions and hadrons faking electrons in the \ee\ sample, are backgrounds that are referred to below as QCD background.
Events with  jets accompanying $W$ bosons (\wpjet) may similarly produce dilepton candidates.  

The expected signal and backgrounds, with the exception of the ones from QCD and \wpjet, are evaluated with simulated samples
and rescaled using the most precise available cross-section predictions, as explained in more detail in section~\ref{sec:expected_s_and_b}.
The total SM prediction is then normalized to the data in an invariant mass interval around the $Z$ peak (70--110~GeV).
In the dielectron channel, the rescaling is done after adding the QCD multijet and \wpjet\ backgrounds evaluated directly from data,
as described in section~\ref{sec:expected_s_and_b}.

%% file: simulation.tex

\section{Simulated samples}
The \zp, \gstar , and LSTC signals,  as well as  the $Z/\gamma^*$ process, are generated with
\pythia\ 6.421~\cite{Sjostrand:2006za} using MRST2007 LO**~\cite{mrst,Sherstnev:2008dm} parton distribution functions (PDFs).
The Minimal \zp\ and \zkk\ signals are obtained by reweighting the large sample of $Z/\gamma^*$ events from \pythia\
with the appropriate ratio of differential cross sections~\cite{Salvioni:2009mt,Salvioni:2009jp,pythia8}. 
\zstar\ and Torsion signals are generated with \comphep~\cite{comphep},
while MadGraph~\cite{MadGraph4} is used for MWT signals;
CTEQ6L1~\cite{Pumplin:2002vw} PDFs are used in both cases.

The diboson processes are generated with \herwig~6.510~\cite{herwig} using MRST2007 LO** PDFs.
The \ttbar\ background is generated with \mcatnlo~4.01~\cite{mcatnlo} using CTEQ66~\cite{Nadolsky:2008} PDFs.
For \ttbar\ events, \jimmy~4.31~\cite{jimmy} is used to describe multiple parton interactions and
\herwig\ to describe the remaining underlying event and parton showers.
Final-state photon radiation is handled by \photos~\cite{fsr_ref}. 
The generated samples are processed through a full ATLAS detector simulation~\cite{atlas:sim} based on GEANT4~\cite{geant}.

\section{Expected signals and backgrounds}
\label{sec:expected_s_and_b}
%
The $Z/\gamma^*$ cross section is calculated at next-to-next-to-leading order (NNLO) in QCD using  
PHOZPR~\cite{Hamberg:1990np} with MSTW2008 NNLO PDFs~\cite{mstw}. 
The ratio of this cross section to the leading-order cross section is used to determine a mass-dependent QCD K-factor,
which is then applied to the results of the leading-order simulation. 
The same QCD K-factor is applied to the \zp , \zkk , Torsion, and LSTC signals. 
Its value  is 0.91 at 2~TeV and slowly increases up to 1.15 at 250~GeV.
A different K-factor is applied to the \gstar\ signal, with 
values that vary between 1.6 and 1.8 depending on the graviton mass and  \kovermb~\cite{Matthews:2009np}, and 
with a value of  1.75 above 750~GeV, consistent with ref.~\cite{ATLAS_grav_diphoton}.
Finally, no QCD K-factor is applied to the leading-order \zstar\ cross section
since the \zstar\ model uses an effective Lagrangian with a different Lorentz structure. 

The \zpssm, \zp(\esix), Torsion states, and techni-mesons interfere minimally with the \dy\ process, and the \zstar\ and \gstar\ do not interfere at all. 
The effect of interference on the resonance line-shape is therefore neglected for all these states.
On the other hand, the interference of the \zkk\ boson with \dy\ is very strong and cannot be neglected \cite{GG,Bella:2010sc}.
The interference effect is also taken into account in the \MM\ framework.

Higher-order electroweak corrections (beyond the photon radiation included in the simulation) 
are calculated using \horace~\cite{horace,CarloniCalame:2007cd}, 
yielding an electroweak K-factor ($K_{{\rm EW}}$) due to virtual heavy gauge boson loops.
Its value at 2~TeV is 0.92 in the dielectron channel and 0.93 in the dimuon channel, and slowly increases up to 1.05 at 250~GeV. 
The electroweak K-factor is applied only to the \dy\ background and not to the expected signals, with the exception of Technicolor and Kaluza-Klein states.
In the case of Technicolor, $K_{{\rm EW}}$ is applied because production proceeds via the \dy\ process.
Since interference is an important feature of the Kaluza-Klein boson model, the electroweak K-factor
is applied to the full amplitude ($\cal{M}$) of the process, including the \zkk\ amplitude: 
                                 $\left|\cal{M}_{\textit{\dy}}+\cal{M}_{\textit{\zkk}}\right|^2 
\longrightarrow K_{{\rm EW}}\times\left|\cal{M}_{\textit{\dy}}+\cal{M}_{\textit{\zkk}}\right|^2$. 
This approximation is conservative.
Although interference is taken into account for Minimal \zp\ bosons,
for consistency with the treatment of the other \zp\ models the electroweak K-factor is applied only to 
the pure \zgstar\ part of the amplitude:
              $\left|\cal{M}_{\textit{\dy}}+\cal{M}_{\textit{\zp}}\right|^2 
\longrightarrow\left|\cal{M}_{\textit{\dy}}+\cal{M}_{\textit{\zp}}\right|^2 + \left(K_{{\rm EW}}-1\right)\times\left|\cal{M}_{\textit{\dy}}\right|^2$.

%
%
For the other backgrounds,
the diboson cross sections are calculated to next-to-leading order (NLO) using {\sc mcfm}~\cite{Campbell:1999mcfm} with an uncertainty of 5\%, 
and the \ttbar\ cross section is predicted at approximate-NNLO, with an uncertainty of $+7.0/-9.6$\%~\cite{Moch:2008qy,Langenfeld:2009tc}.

At very high masses, the statistical significance of the diboson and \ttbar\ simulated samples becomes insufficient.
Therefore their invariant mass distribution is fitted
to the functional form $y(x)= p_{1}\cdot x^{p_{2}+p_{3}\log{x}}$
which is then used to extrapolate the \ttbar\ background above 0.8~TeV and the diboson background above 1.5~TeV.

%% file: data_driven.tex



The QCD multijet and \wpjet\ backgrounds in the \ee\ sample are estimated primarily from data using several techniques.

First, a ``reversed electron identification'' technique~\cite{Aad:2011xp} is used, in which
only the QCD multijet background is estimated from data, while the \wpjet\ component comes from the Monte Carlo simulation.
Events with both electron candidates failing one of the \medium\ identification criteria 
are used to determine the shape of the QCD multijet background \mee\ distribution. 
The chosen criterion, the difference in $\eta$ between the cluster and the track, does not affect kinematic distributions.
The small contamination from non-QCD processes, located mainly beneath the $Z$ peak, 
is subtracted  
using MC samples of the other backgrounds (\dy, \ttbar , diboson and \wpjet).
The shape of these backgrounds is obtained by summing their contributions 
according to the most precise available cross-section predictions.
The mass distributions of the QCD and non-QCD backgrounds are fitted for their relative contributions in the 70~GeV $< \mee < 200$~GeV range.   
The QCD multijet background shape is fitted in the 110--800~GeV range
using the functional form $y(x)= p_{1}\cdot x^{p_{2}+p_{3}\log{x}}$ and is extrapolated beyond 800~GeV.
The systematic uncertainty 
includes the uncertainty from the relative normalization procedure, namely the QCD multijet fraction from the first fit, the uncertainty
from the choice of the reversed selection and the uncertainty from the range of the second fit.

A second independent data-driven method is used to obtain an estimate of the QCD multijet and \wpjet\ backgrounds together.
It uses fake rates computed from jet-enriched samples, obtained from jet triggers or from the signal trigger. 
The fake factors are defined as the probability for a jet to pass the ``tight'' (T) selection, that is all selection criteria, 
if it passes the ``loose'' (L) selection, that is the same reverse identification selection as in the reverse electron identification method. 
The fake factors depend only slightly on transverse momentum and more importantly on pseudorapidity.
Since isolation is applied only to the leading electron, two different fake factors are needed: 
\flead, applied to the leading electron, and \fsubl, applied to the subleading electron.
The QCD multijet and $\ell$+jets backgrounds are estimated by selecting events with candidate pairs 
having each electron identified either as ``tight'' 
or ``loose''.
The final estimate is then

\begin{equation*}
N_{\ell+{\rm jets} \ \& \, {\rm QCD}} =  \fsubl N_{\rm TL} + \flead N_{\rm LT} - \flead \fsubl N_{\rm LL}. 
\end{equation*}
The same functional form as before is finally used to fit this estimate between 140~GeV and 850~GeV and extrapolate it above
this energy range.
The systematic uncertainty includes the uncertainty from the $\eta$ or \pt\ dependence of the fake factors
and the uncertainty from the range of the fit.

All methods yield consistent results and the final estimate is given by the mean of the central values.
The uncertainty, conservatively assigned to be the maximum of the largest of up and down deviations of each method, 
is 33\% at $\mee=200$~GeV and grows to about 110\% at 2~TeV.

In the dimuon channel, the QCD multijet background is estimated in data from a sample
of non-isolated dimuon events. The  \wpjet\  background  is evaluated using simulated samples. 
Both backgrounds are found to be negligible in the dimuon channel after the isolation selection is applied.

%% file: systematics.tex

The systematic uncertainties in this analysis are reduced by the fact that the backgrounds are normalized to the data in the region of the $Z$ peak. 
This procedure makes the analysis insensitive to the error on the measurement of the integrated luminosity as well as other mass-independent systematic uncertainties. 
Instead, a constant systematic uncertainty of 5\%, due to the uncertainty on the $Z/\gamma^*$ cross section in the normalization region, is assigned to the signal expectation. 

The mass-dependent systematic uncertainties include theoretical effects due to the PDFs, QCD and electroweak corrections, 
as well as experimental effects, namely efficiency and resolution. 
These uncertainties are correlated across all bins in the search region. 
In addition, there is an uncertainty on the QCD and \wpjet\ backgrounds affecting the dielectron channel. 
The theoretical uncertainties are applied to the background expectation only.
The experimental uncertainties are assumed to be correlated between signal and all types of backgrounds.
All systematic uncertainties quoted below refer to narrow resonances with dilepton masses of  2~TeV.
All systematic uncertainties estimated to have an impact $\leq 3\%$  on the expected number of events are neglected. 

The combined uncertainty on the PDFs, strong coupling \alphas, and renormalization/facto-rization
scale variations
is 20\%, the largest contribution being the uncertainty on the PDFs. 
The \alphas\ and PDF uncertainties are evaluated using the MSTW2008NNLO 
eigenvector PDF sets and the PDF sets corresponding to variations of \alphas, at the 90\% confidence level (CL).
The spread of the variations covers the difference between the central values obtained with the CTEQ and MSTW PDF sets.  
The scale uncertainties are estimated by varying the renormalization (\mur)
and factorization (\muf) scales independently up and down by a factor of 
two, but with the constraint $0.5\le \muf/\mur\le 2$
to avoid large logarithmic corrections. 
The resulting maximum variations are taken as the uncertainties.
In addition, a systematic uncertainty of 4.5\% is attributed to electroweak corrections~\cite{Aad:2011xp} for both channels.
This contribution includes 
the difference in the
electroweak scheme definition between \pythia\ and \horace, and higher order electroweak and $\mathcal{O}(\alpha\alphas)$ corrections.

In the dielectron channel, the largest experimental systematic uncertainty is due to  
the estimate of the QCD multijet and \wpjet\ backgrounds, 
which translates into a systematic uncertainty on the total background of 26\% at 2~TeV.   
Other experimental systematic uncertainties in the dielectron channel include 
uncertainties due to the extrapolation of the \ttbar\ and diboson backgrounds, which are significant only above 2~TeV and 
uncertainties due to the electron reconstruction and identification efficiency at high \et , which are estimated to be less than 3\% for electron pairs. The uncertainties on the calorimeter energy calibration are estimated to be between 0.5\% and 1.5\%,
depending on transverse energy and pseudorapidity and have a negligible effect on the event yield as do the uncertainties on the corrections applied to the simulation to reproduce the calorimeter resolution at high energy.

In the dimuon channel, the combined uncertainty on the trigger and reconstruction efficiency for muon pairs is estimated to be 6\% at 2~TeV.  
This uncertainty is dominated by a conservative estimate of the impact from large energy 
loss due to muon bremsstrahlung in the calorimeter, which may interfere with reconstruction in the muon spectrometer.
In addition, the uncertainty on the resolution due to residual misalignments in the muon spectrometer propagates to a change in the observed
width of the signal line-shape; however, its effect on the final result 
has a negligible impact for final states which do not interfere strongly with \dy .  
Finally, the muon momentum scale is calibrated with a statistical precision of 0.1\% using the  $Z\to\ll$ mass peak. 
As with the dielectron channel, the momentum calibration uncertainty has negligible impact in the dimuon channel search. 

A summary of all systematic uncertainties common to all final states investigated in this search is shown in table~\ref{tab:systematicSummary}.  
Additional systematic uncertainties that apply to strongly interfering states (such as \zkk) are discussed later. 
\begin{table}[!t]
\caption{
Summary of systematic uncertainties on the expected numbers of events at $\mll=2$~TeV. 
NA~indicates that the uncertainty is not applicable, and ``-'' denotes a negligible entry.
}
\label{tab:systematicSummary}
\centering
\addtolength{\tabcolsep}{+1pt}
\small
\begin{tabular}{l|cc|cc}
\hline
\hline
Source          			& \multicolumn{2}{c|}{Dielectrons}    & \multicolumn{2}{c}{Dimuons} \\
                	    			& Signal & Background                                 &  Signal & Background \\
\hline
Normalization			& 5\%     & NA			& 5\%   & NA \\
PDF/$\alpha_{s}$ /scale	        & NA      & 20\%	        & NA    & 20\% \\
Electroweak corrections		& NA      & 4.5\%		& NA    & 4.5\% \\
Efficiency			& -       & -			& 6\%   & 6\% \\
\wpjet\ and QCD background 	& NA      & 26\%		& NA    & - \\
\hline		       		         				      			 
 Total				&  5\%    & 34\%		& 8\%   & 21\%\\
\hline
\hline
\end{tabular}

\end{table}

%% file: results.tex

\section{Data-SM expectation comparison}

\begin{figure}[tbp]
  \includegraphics[width=0.9\textwidth]{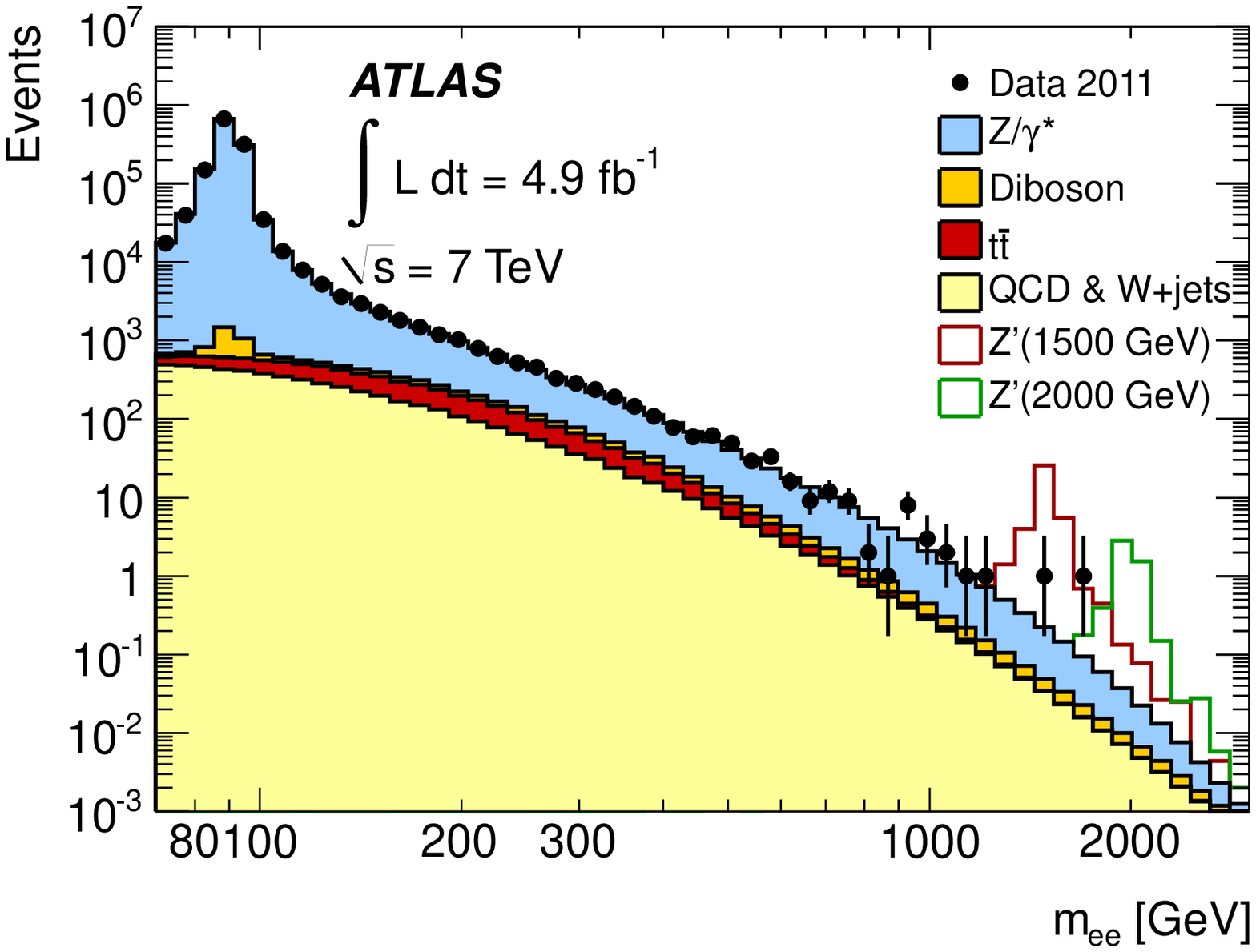}
  \includegraphics[width=0.9\textwidth]{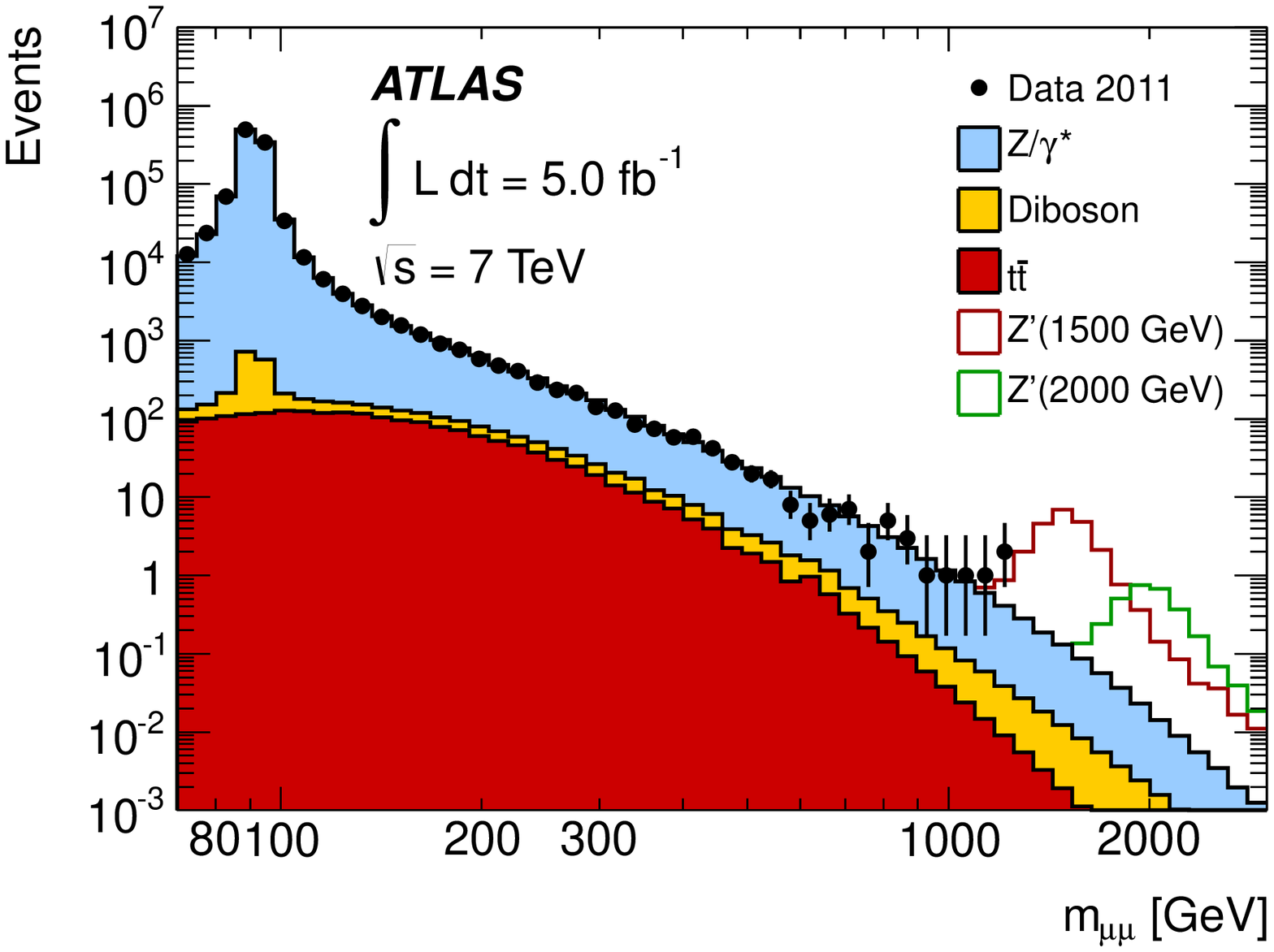}  
  \caption{Dielectron (top) and dimuon (bottom) invariant mass (\mll) distributions after final selection, 
           compared with the stacked sum of all expected backgrounds, with two
example \zpssm\ signals overlaid. 
           The bin width is constant in $\log \mll$.}
  \label{fig:mll}
\end{figure}

Figure~\ref{fig:mll}  shows  the invariant mass (\mll) distribution for the dielectron (top) and dimuon (bottom) final states after final 
selection. 
The bin width of the histograms is constant in $\log \mll$, chosen such that a possible signal peak spans multiple bins and the templates are smooth.
Figure~\ref{fig:mll} also displays the expected \zpssm\ signal for two mass hypotheses. 
Tables~\ref{tab:backgroundTableEE} and~\ref{tab:backgroundTableMuon}
 show the number of data events and the estimated backgrounds in bins of reconstructed dielectron and dimuon invariant mass above 110~GeV.
The number of observed events in the normalization region, from 70 to 110 GeV, is 1,236,646 in the dielectron channel and 985,180 in the dimuon channel.
The dilepton invariant mass distributions are well described by the Standard Model.

\begin{table}[tbp]
\caption{Expected and observed number of events in the dielectron channel.  
         The errors quoted include both statistical and systematic uncertainties.}
\label{tab:backgroundTableEE}  
\begin{center}
\begin{tabular}{lccccc}
\hline\hline
\mepem [GeV]    &   110--200          &   200--400        &   400--800      &   800--1200       &  1200--3000\\
\hline
\zgstar         & $ 26700 \pm 1100 $  & $ 2960 \pm 120 $  & $ 265 \pm 13 $  & $ 12.1 \pm 0.9 $  & $ 1.47 \pm 0.18 $ \\
\ttbar          & $ 1300 \pm 120 $  & $ 410 \pm 40 $  & $ 26.5 \pm 2.8 $  & $ 0.41 \pm 0.17 $  & $ 0.034 \pm 0.034 $ \\
Diboson         & $ 415 \pm 21 $  & $ 146 \pm 8 $  & $ 16.2 \pm 0.9 $  & $ 0.88 \pm 0.05 $  & $ 0.101 \pm 0.011 $ \\
QCD and \wpjet  & $ 1900 \pm 600 $  & $ 510 \pm 200 $  & $ 50 \pm 31 $  & $ 2.0 \pm 1.8 $  & $ 0.26 \pm 0.31 $ \\
\hline
Total           & $ 30300 \pm 1300 $  & $ 4030 \pm 240 $  & $ 357 \pm 34 $  & $ 15.4 \pm 2.0 $  & $ 1.86 \pm 0.35 $ \\
\hline
Data            & $ 29816$ & $  4026$ & $   358$ & $    17$ & $     3$\\
\hline\hline
\end{tabular}
\end{center}
\end{table}
\begin{table}[tbp]
\caption{Expected and observed number of events in the  dimuon channel. 
         The errors quoted include both statistical and systematic uncertainties.}
\label{tab:backgroundTableMuon}
\begin{center}
\begin{tabular}{lccccc}
\hline\hline
\mmumu\ [GeV] 	   &   110--200 	&   200--400 	&   400--800 	&   800--1200 	&  1200--3000\\ 
\hline
\zgstar & $ 21200 \pm 1200 $ & $ 2090 \pm 230 $ & $ 173 \pm 15 $ & $ 7.7 \pm 0.8 $ & $ 0.98 \pm 0.16 $\\  
\ttbar  & $ 900 \pm 100 $ & $ 270 \pm 50 $ & $ 18 \pm 11 $ & $ 0.32 \pm 0.07 $ & $ 0.019 \pm 0.007 $\\  
Diboson & $ 289 \pm 32 $ & $ 97 \pm 24 $ & $ 11.8 \pm 2.7 $ & $ 0.59 \pm 0.26 $ & $ 0.087 \pm 0.016 $\\  
\hline
Total   & $ 22400 \pm 1200 $ & $ 2460 \pm 240 $ & $ 203 \pm 19 $ & $ 8.7 \pm 0.9 $ & $ 1.09 \pm 0.16 $\\  
\hline
Data    & $ 21945$ & $  2294$ & $   197$ & $    10$ & $     2 $\\ 
\hline\hline
\end{tabular}
  \end{center}
\end{table}

The data are compared to the Monte Carlo simulation in the search region 0.13~TeV$<\mll <3.0$~TeV.
%
The agreement is first studied by computing the significance of the difference in each mass bin,
with statistical and systematic uncertainties taken into account.
The largest positive local significance is about $2\sigma$ in the dielectron channel and about $1\sigma$ in the dimuon channel,
and the largest negative local significance is $-2\sigma$ in both channels.

The comparison is then performed by means of templates~\cite{Aad:2011xp, CDF:Zpmumu2fb}.
The templates provide the expected yield of events ($\bar{n}$) in each \mll\ bin. 
When neglecting interference, $\bar{n}$ is given by $\bar{n} = n_X(\lambda , {\pmb\nu}) + n_{\dy} ({\pmb\nu}) + \nobg ({\pmb\nu})$,
where $n_{X}$ represent the number of events produced by the decay of a new resonance $X$ 
($X=\zp ,\zstar , \gstar , \ts, \rhot/\omegat , \Ronetwo$, where \rhot/\omegat\ and \Ronetwo\ are techni-mesons, see below);
$n_{\dy}$ and \nobg\ are the number of \dy\ (Drell-Yan) and other backgrounds events, respectively.
The symbol $\lambda$ represents the parameter of interest of the model, 
and ${\pmb\nu}$ is the set of Gaussian-distributed nuisance parameters incorporating the systematic uncertainties.
When including  the effects of interference, $\bar{n} = n_{X+\dy}(\lambda , {\pmb\nu}) + \nobg ({\pmb\nu})$,
where $n_{X+\dy}$ is the number of signal plus \zgstar\ events and $X$ can be \zkk\ or a Minimal \zp\ boson.
Signal templates provide the expected line-shape of the dilepton resonances.


The significance of a signal is summarized by a \pval, the probability of observing
a signal-like excess at least as extreme as the one observed in data, assuming the null hypothesis.
The outcome of the search is ranked using a log-likelihood ratio (LLR),
with the likelihood function defined as the product of the Poisson probabilities over all mass bins in the search region,
using a \zpssm\ template.
Explicitly:

\begin{equation*}
{\rm LLR} = -2\ {\rm ln}\ \frac{\mathcal{L} ({\rm data}\ |\ \hat{n}_{\zp}, \hat{M}_{\zp}, \hat{\pmb\nu} ) }{\mathcal{L} ({\rm data}\ |\ (\hat{n}_{\zp} = 0), \hat{\hat{\pmb\nu}} ) }
\end{equation*}
where $\hat{n}_{\zp}$, $\hat{M}_{\zp}$, $\hat{\pmb\nu}$ and $\hat{\hat{\pmb\nu}}$ 
are respectively the best-fit values for the \zp~normalization, \zp~mass and nuisance parameters,
which maximize the likelihood~$\mathcal{L}$ given the data,
assuming in the numerator that a \zp\ signal is present and in the denominator that no signal is present.
The LLR is scanned as a function of \zp\ cross section and \mzp\ over the full considered mass range.
The observed \pval\ for the dielectron and dimuon samples is \pvalele\ and \pvalmuon, respectively. 
For the combination of both channels, the observed \pval\ is \pvalcomb.

%% file: simple_limits.tex

In the absence of a signal, upper limits on the number of events produced by the decay of a new resonance 
are determined at the 95\% Confidence Level (CL).

The limit on the number of signal events is converted into a limit on the ratio of cross section times branching fraction $\xbr (X\to\ll ) / \xbr (Z \to \ll)$ by dividing by the observed number of \z\ boson events and the ratio of corresponding acceptances. 
This ratio of \xbr\ is then converted into a limit on $\xbr (X\to\ll )$ by multiplying it by the theoretical value of $\xbr (Z \to \ll)$.
Because of the strong destructive interference between Kaluza-Klein
bosons and \dy, limits are set on the coupling strength of the resonance to the fermions instead of the cross section times branching ratio. The same is done for the class of \MM, where the coupling strength \gammap\ is one of the two parameters defining the model.

The same Bayesian approach~\cite{bayesianMethod} is used in all cases,
with a flat prior probability distribution for
the signal cross section times branching fraction (\xbr), when neglecting interference.
When including the effects of interference, the prior is flat for the coupling strength to the second or to the fourth power.
The most likely number of signal events, and the corresponding confidence intervals, 
are determined from a likelihood function defined as the product of the Poisson probabilities over all mass bins in the search region,
using the appropriate signal templates.
The nuisance parameters are integrated out.

Most of the dilepton resonances searched for in this analysis are narrow compared to the detector resolution.
The effect of width variations on the resonance line-shape is neglected for the \esix\ analysis. 
On the other hand, the dependence of the width on the coupling strength is taken into account in 
the MWT, Torsion and \gstar\ analyses by using several templates for a given pole mass 
in which various values of the couplings are selected. 
Signal templates include the acceptance times efficiency of the signal, at a given pole mass $M_X$, over the full search region.
The product \acc\ is different for each model due to different angular distributions, boosts, and line-shapes.

The expected exclusion limits are determined using simulated pseudo-experiments with only Standard Model processes 
by evaluating the 95\% CL upper limits for each pseudo-experiment for each fixed value of the resonance pole mass $M_{X}$.
The median of the distribution of limits is chosen to represent the expected limit. The ensemble of limits
 is also used to find the  68\% and 95\% envelopes of the expected limits as a function of $M_{X}$.

The combination of the dielectron and dimuon channels is performed under the 
assumption of  lepton universality  by defining the likelihood function in terms of 
the total number of signal events produced in both channels. 
For each source of uncertainty, the correlations across bins, as well as the correlations between signal and background, are taken into account.

\section{Limits on spin-1 SSM and \esix\ \zp\ bosons } 

Due to mixing between the $U(1)_\chi$ and $U(1)_\psi$ groups, 
 in the \esix\ models  the lightest new boson is a linear combination
 of the \zpchi\ and \zppsi\ bosons depending on the mixing angle \te6. 
For six specific values of this mixing angle, the 
diboson resonance is named $Z'_{\psi}$,  \zpN,  \zpeta, \zpI,  \zpsq , and $\zp _{\chi}$.
The corresponding mixing angle values are displayed in table~\ref{tab:e6angleDef}.
Like the SSM, these models prescribe the couplings of the \zp\ boson to the SM fermions. 
The expected intrinsic width of the \zp\ boson in the  \esix\ models  
is predicted to be between 0.5\% and 1.3\%~\cite{Dittmar:2003ir,Accomando:2010fz} 
of its mass,  while in the SSM the intrinsic width is predicted to be about $3$\%. 
\begin{table}[tbp]
\caption{
Mixing angle values for the \esix\ models considered.
}
\label{tab:e6angleDef}
\begin{center}
\begin{tabular}{l|cccccc}
\hline
\hline
Model        & \zppsi & \zpN          & \zpeta       & \zpI          & \zpsq          & \zpchi  \\
\hline			 	 	 	 	  	 
$\sin\te6$   & 0      & $-1/4$        & $\sqrt{3/8}$ & $\sqrt{5/8}$  & $3\sqrt{6}/8$  & 1 \\
$\cos\te6$   & 1      & $\sqrt{15}/4$ & $\sqrt{5/8}$ & $-\sqrt{3/8}$ & $-\sqrt{10}/8$ & 0 \\
\hline
\hline
\end{tabular}
\end{center}
\end{table}

\begin{figure}[!t]
  \centering
  \includegraphics[width=0.49\columnwidth]{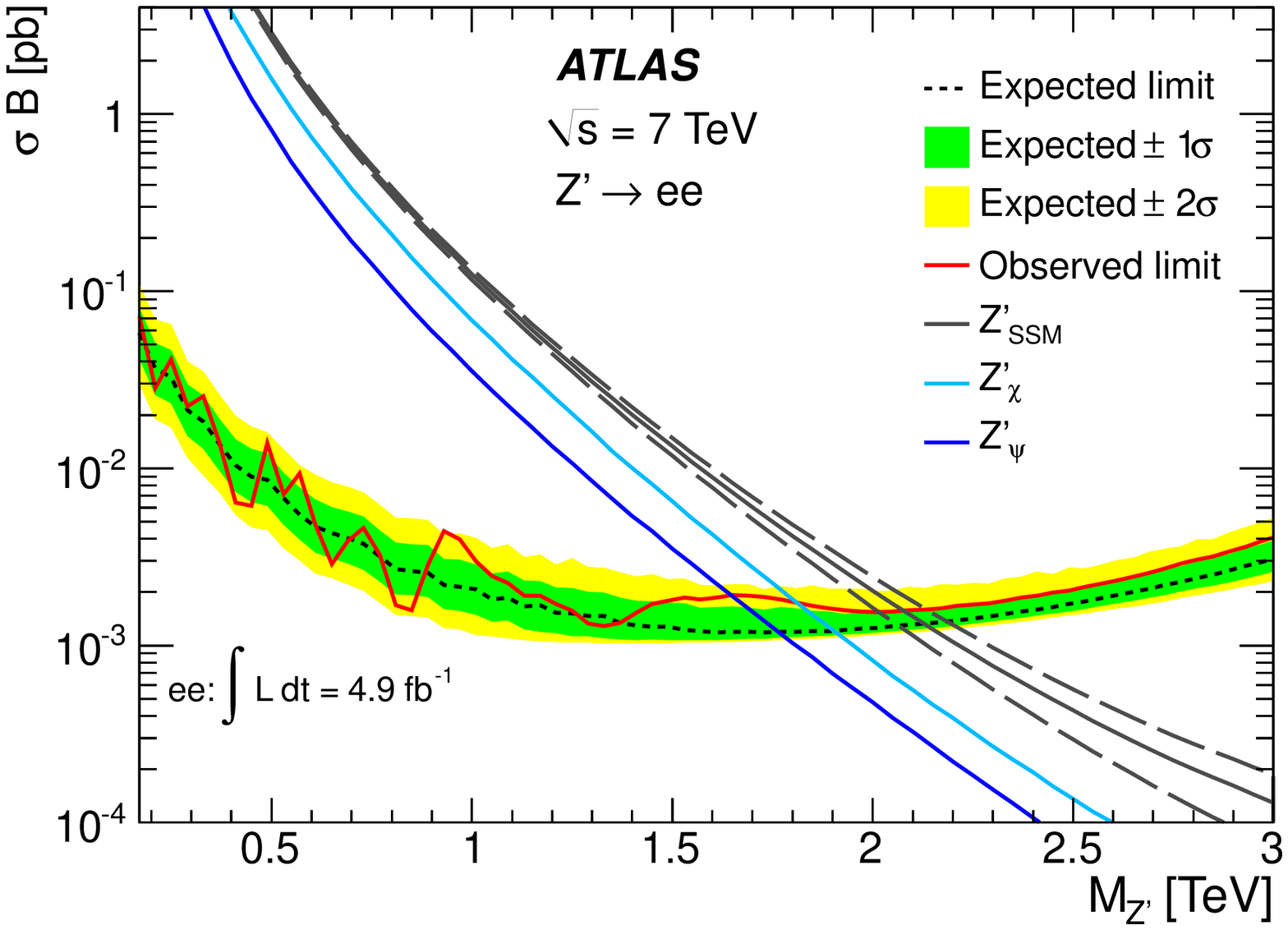}
  \includegraphics[width=0.49\columnwidth]{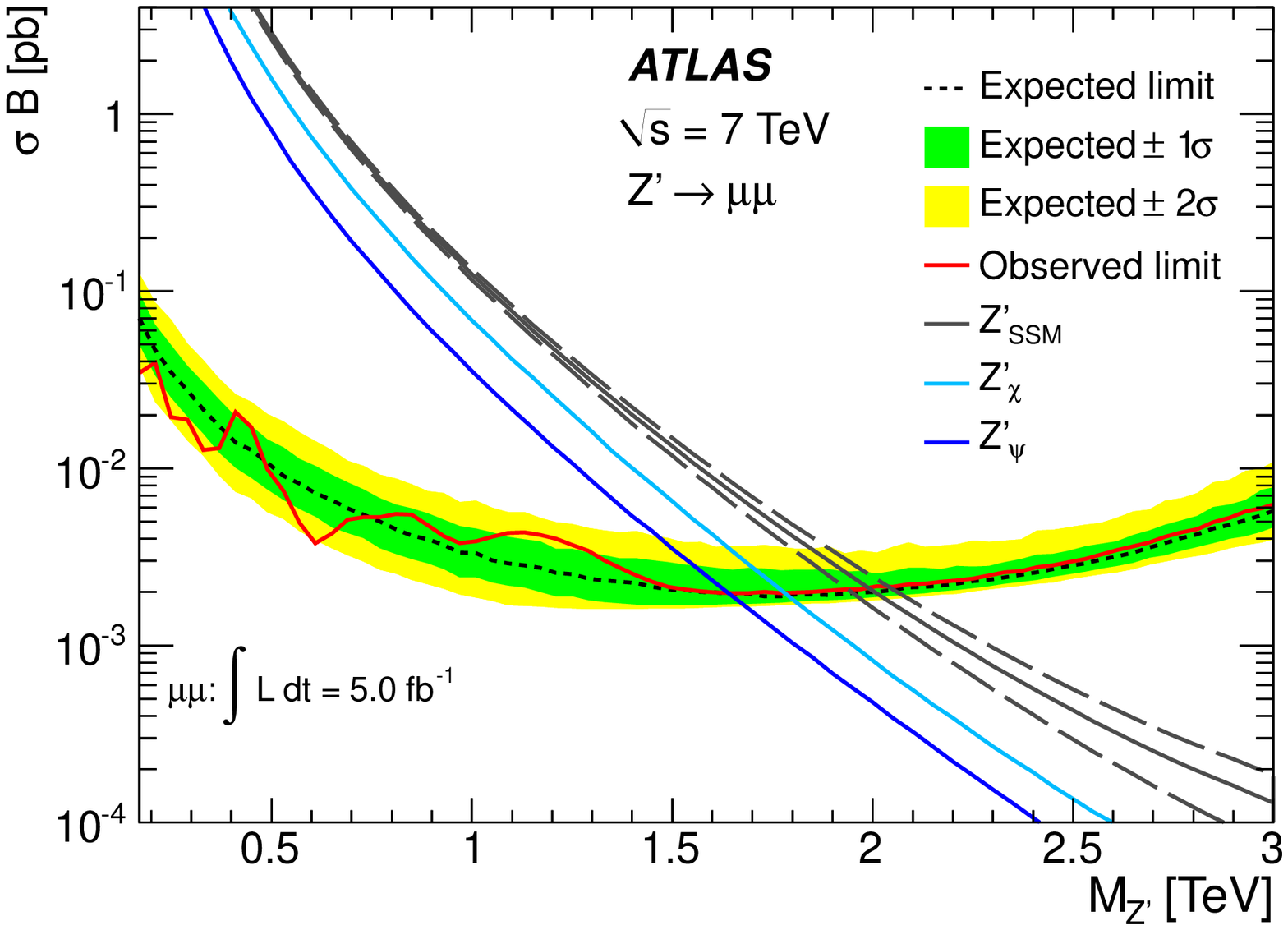}
   \caption{Expected and observed 95\% CL limits on \xbr\
        and expected \xbr\ for \zpssm\ production and the two \esix-motivated \zp\ models with lowest and highest \xbr\ 
        for the dielectron (left), and the dimuon (right) selections.
        The dashed lines around the \zpssm\ theory curve represent the theoretical uncertainty, which is similar for the other theory curves.
}
  \label{fig:Zplimit_res}
\end{figure}

\begin{figure}[tbp]
  \centering
  \includegraphics[width=0.7\columnwidth]{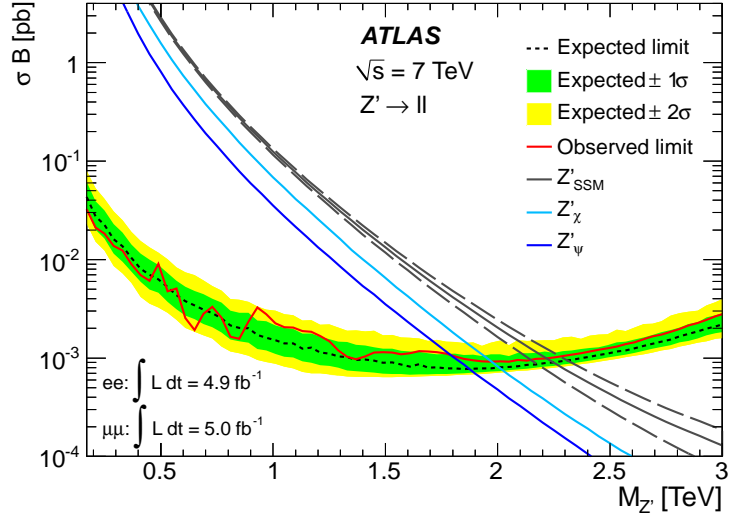}
  \caption{Expected and observed 95\% CL limits on \xbr\
           and expected \xbr\ for \zpssm\ production and the two \esix-motivated \zp\ models with lowest and highest \xbr\ 
           for the combination of the dielectron and dimuon channels.
            The dashed lines around the \zpssm\ theory curve represent the theoretical uncertainty, which is similar for the other theory curves.
}
  \label{fig:combinedlimit_res}
\end{figure}

Figure~\ref{fig:Zplimit_res} shows the 95\% CL observed and expected exclusion limits on $\xbr (\zp \to \ee)$ and  $\xbr (\zp \to \mumu)$
obtained with \zpssm\ templates.
It also shows the theoretical cross section times branching fraction for the \zpssm\ and for the lowest and highest \xbr\ of \esix-motivated \zp\ models.
The combination of  the dielectron and dimuon channels is shown in figure~\ref{fig:combinedlimit_res}. 
The rise of the \xbr\ limit at high invariant mass is due mainly to the fast fall of the parton luminosity at high momentum transfer which enhances the low-mass tail, causing a distortion in the resonance peak shape. 

The 95\% CL \xbr\ limit is used to set mass limits for each of the models considered. 
The limits obtained for the \zpssm\ are displayed in table~\ref{tab:limits}.
The combined observed (expected) mass limit for the \zpssm\ is 
\LimitCombined\ (\LimitCombinedExpected)~TeV. 
The combined mass limits on \esix-motivated \zp\ are given in table~\ref{e6massLimits}.
\begin{table}[tbp]
\caption{
The observed and expected 95\% CL lower limits on the mass of the \zpssm\ boson for the \ee\ and \mumu\ 
channels separately and for their combination.
}
\label{tab:limits}
\begin{center}
\begin{tabular}{l|ccc}
\hline
\hline
                     & $\zpssm \to \ee$       & $\zpssm \to \mumu$ & $\zpssm \to \ll$ \\
\hline			 	 	 	 	  	 
Observed limit [TeV] & \LimitElectron         & \LimitMuon         & \LimitCombined  \\
Expected limit [TeV] & \LimitElectronExpected & \LimitMuonExpected & \LimitCombinedExpected \\
\hline
\hline
\end{tabular}
\end{center}
\end{table}
\begin{table}[tbp]
\caption{
The observed and expected 95\% CL lower limits on the masses of \esix-motivated \zp\ bosons.
Both lepton channels are combined.
}
\label{e6massLimits}
\begin{center}
\begin{tabular}{l|cccccc}
\hline
\hline
Model                     & \zppsi & \zpN  & \zpeta & \zpI  & \zpsq & \zpchi  \\
\hline			 	 	 	 	  	 
Observed limit [TeV]&\LimitCombinedPsi        &\LimitCombinedN        &\LimitCombinedEta        &\LimitCombinedI        &\LimitCombinedSq        &\LimitCombinedChi \\
Expected limit [TeV]&\LimitCombinedExpectedPsi&\LimitCombinedExpectedN&\LimitCombinedExpectedEta&\LimitCombinedExpectedI&\LimitCombinedExpectedSq&\LimitCombinedExpectedChi   \\
\hline
\hline
\end{tabular}
\end{center}
\end{table}

\section{Limits on spin-1 \zstar\ bosons } 

A model with quark-lepton universality is adopted~\cite{wzstar_refmod,wzstar_refmod2}
to fix the coupling strength of the \zstar\ boson to fermions.
The gauge coupling is chosen to be the same as in the SM SU(2) group,
and the scale of the new physics is proportional to the mass of the new heavy bosons. 
The parameters of the model are fixed by requiring 
that the total and partial decay widths of \wstar, the charged partner of \zstar, 
be the same as  those of the \wpssm\ boson with the same mass. The width of the \zstar\ is then 3.4\% of its mass.
As a result of the tensor form of the coupling, 
the \zstar\ does not interfere with \dy, and 
the  angular distribution of its decay to dileptons
is different from that of a \zp\ boson. 

\begin{figure}[tbp]
  \centering
\includegraphics[width=0.7\textwidth]{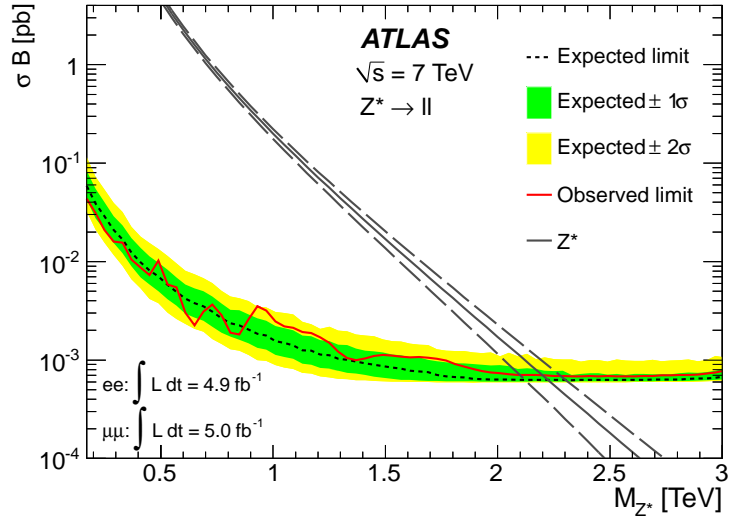}
  \caption{Expected and observed 95\% CL limits on \xbr\ and expected \xbr\ for \zstar\ boson production for the combination of 
           dielectron and dimuon channels. The dashed lines around the \zstar\ theory curve represent the theoretical uncertainty.
}
  \label{fig:combinedlimit_Zs}
\end{figure}

Figure~\ref{fig:combinedlimit_Zs} shows the 95\% CL observed and expected exclusion limits on $\xbr (\zstar \to \ll)$ 
as well as the cross section times branching fraction expected from theory.
The corresponding 95\% CL limits on the mass of the \zstar\ boson are shown in table~\ref{tab:limits_Zs}.
\begin{table}[tbp]
\caption{
The observed and expected 95\% CL lower limits on the mass of the \zstar\ boson for the \ee\ and \mumu\ channels separately and for their combination. 
}
\label{tab:limits_Zs}
\begin{center}
\begin{tabular}{l|ccc}
\hline
\hline
                     & $\zstar \to \ee$         & $\zstar \to \mumu$   & $\zstar \to \ll$ \\
\hline			 	 	 	 	  	 
Observed limit [TeV] & \LimitElectronZs         & \LimitMuonZs         & \LimitCombinedZs  \\
Expected limit [TeV] & \LimitElectronExpectedZs & \LimitMuonExpectedZs & \LimitCombinedExpectedZs \\
\hline
\hline
\end{tabular}
\end{center}
\end{table}

\section{Limits on spin-2 Randall-Sundrum gravitons} 

\begin{figure}[tbp]
  \centering
  \includegraphics[width=0.7\textwidth]{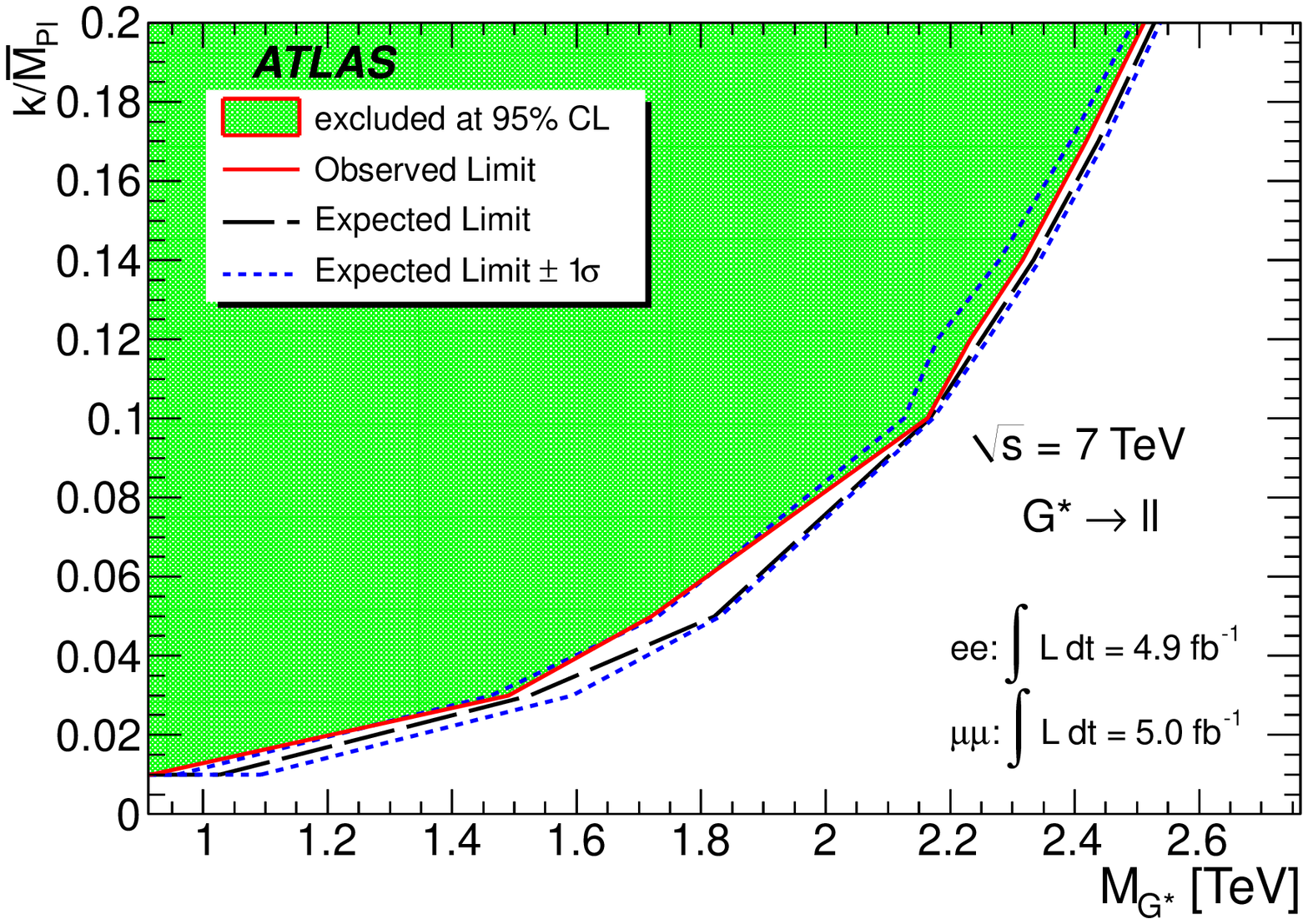}
  \caption{Exclusion regions in the plane of \kovermb\ versus graviton mass for the combination of dielectron and dimuon channels.
           The region above the curve is excluded at 95\% CL. 
	  }
  \label{fig:gstar_2Dlim_comb}
\end{figure}

The phenomenology of the RS model used in this work can be described in terms of the mass of the graviton 
and the ratio \kovermb . 
The expected intrinsic width of the \gstar\ is proportional to $(\kovermb)^2$, and is 1.4\% for $\kovermb =0.1$.
Limits at the 95\% CL on
$\xbr (\gstar \to \ell^+\ell^-)$ are computed
assuming two values of \kovermb:
0.1 and 0.2. 
These limits are then compared to the theoretical cross section times branching fraction assuming eight different 
values of \kovermb\ between 0.01 and 0.2. 
The \xbr\ limits obtained with $\kovermb=0.1$ 
are used for \kovermb\ hypotheses below or equal to 0.1, 
while those with  $\kovermb=0.2$ are used for \kovermb\ hypotheses larger than 0.1 and below or equal to 0.2.
Limits at the 95\% CL on the graviton mass are derived from this comparison for each \kovermb\ hypothesis and are shown 
in table~\ref{tab:limits_Gs} for $\kovermb=0.1$,
and in table~\ref{tab:combinedLimitsG} and figure~\ref{fig:gstar_2Dlim_comb} for the combined dilepton channel for all values of \kovermb.

\begin{table}[tbp]
\caption{The observed and expected 95\% CL lower limits on the mass of the \gstar\ with a coupling of \kovermb $=0.1$ for the \ee\ and \mumu\ channels separately and for their combination.
}
\label{tab:limits_Gs}
\begin{center}
\begin{tabular}{l|ccc}
\hline
\hline
                     & $\gstar \to \ee$        & $\gstar \to \mumu$  & $\gstar \to \ll$ \\
\hline			 	 	 	 	  	 
Observed limit [TeV] & \LimitElectronG         & \LimitMuonG         & \LimitCombinedG  \\
Expected limit [TeV] & \LimitElectronExpectedG & \LimitMuonExpectedG & \LimitCombinedExpectedG \\
\hline
\hline
\end{tabular}
\end{center}
\end{table}
\begin{table}[tbp]
\caption{
The observed and expected 95\% CL lower limits on the mass of the \gstar\ with varying coupling \kovermb. 
Both lepton channels are combined.
}
\label{tab:combinedLimitsG}
\begin{center}
\begin{tabular}{l|cccccccc}
\hline
\hline
\kovermb            &  0.01 & 0.03  & 0.05 & 0.1 & 0.12 & 0.14 & 0.17 & 0.2\\
\hline			 	 	 	 	  	 
Observed limit [TeV] & \LimitCombinedGOne & \LimitCombinedGThree  & \LimitCombinedGFive  & \limitGraviton & \LimCombGTwelve & \LimCombGFourteen & \LimCombGSeventeen & \LimCombGTwenty \\
Expected limit [TeV] & \LimitCombinedGExpOne & \LimitCombinedGExpThree  & \LimitCombinedGExpFive  & \limitGravitonExp & \LimCombGExpTwelve & \LimCombGExpFourteen & \LimCombGExpSeventeen & \LimCombGExpTwenty \\
\hline
\hline
\end{tabular}
\end{center}
\end{table}

\clearpage

\section{Limits on Torsion models} 
The Torsion heavy state (TS) can be treated as a fundamental propagating field characterized by its mass, \mts , 
and the couplings between TS  and fermions. These couplings are assumed to be universal at the Planck scale
and remain so at the TeV scale for all fermions except the top quark~\cite{Belyaev:2007fn}. 
Therefore the phenomenology of Torsion decays to dilepton states 
can be described in terms of two parameters: the TS mass  and one coupling (\etats ).
Since \etats\ can {\em a priori} take any value between 0 and 1, the intrinsic width could be very large. 
The interference effects with \dy\ are negligible. 

\begin{figure}[tbp]
 \centering
  \includegraphics[width=0.7\columnwidth]{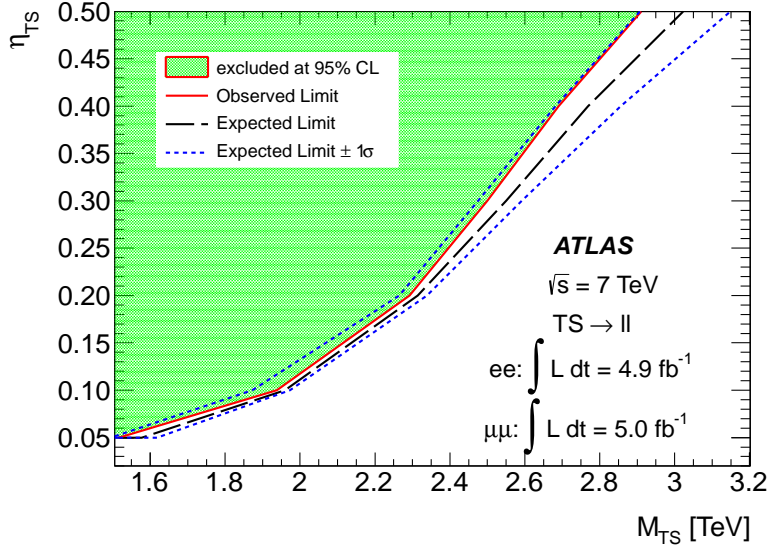}
  \caption{Exclusion regions in the plane of \etats\ versus Torsion mass for the combination of dielectron and dimuon channels.
           The region above the curve is excluded at 95\% CL. 
}
\label{fig:TS_2D_comb}
\end{figure}

Limits are computed on $\xbr (\ts\to\ll)$ 
for five values of \etats\ in the range 0.1--0.5. 
Limits on \xbr\ are then translated into limits on \mts\ 
in the same way as above for the RS graviton, by comparing them to the theoretical \xbr\
as a function of \mts\ 
for each value of \etats .
Additionally, the \xbr\ limits obtained for $\etats=0.1$ are used to set mass limits for $\etats=0.05$,
which is conservative because the TS width is smaller for $\etats=0.05$.
The resulting exclusion region in the (\mts, \etats) plane is displayed in 
figure~\ref{fig:TS_2D_comb} and table~\ref{tab:combinedLimitsTS} for the combined dielectron and dimuon channels.
The limits on \mts\ obtained in each channel for $\etats=0.2$ are shown in table~\ref{tab:limits_Ts}.
\begin{table}[tbp]
\caption{
The observed and expected 95\% CL lower limits on the mass of Torsion heavy states with varying coupling \etats. 
Both lepton channels are combined.
}
\label{tab:combinedLimitsTS}
\begin{center}
\begin{tabular}{l|cccccc}
\hline
\hline
\etats               &  0.05 & 0.1  & 0.2 & 0.3 & 0.4 & 0.5 \\
\hline			 	 	 	 	  	 
Observed limit [TeV] & \LimCombTSHalf    &\LimCombTSOne   &\LimCombTSTwo   &\LimCombTSThree   &\LimCombTSFour   &\LimCombTSFive\\
Expected limit [TeV] & \LimCombExpTSHalf &\LimCombExpTSOne&\LimCombExpTSTwo&\LimCombExpTSThree&\LimCombExpTSFour&\LimCombExpTSFive\\
\hline
\hline
\end{tabular}
\end{center}
\end{table}
\begin{table}[tbp]
\caption{
The observed and expected 95\% CL lower limits on the mass of Torsion heavy states with a coupling of $\etats=0.2$ for the \ee\ and \mumu\ channels separately and for their combination.
}
\label{tab:limits_Ts}
\begin{center}
\begin{tabular}{l|ccc}
\hline
\hline
                     & $\ts \to \ee$         & $\ts \to \mumu$   & $\ts \to \ll$ \\
\hline			 	 	 	 	  	 
Observed limit [TeV] & \LimEleTSTwo          & \LimMuTSTwo     & \LimCombTSTwo  \\
Expected limit [TeV] & \LimEleExpTSTwo       & \LimMuExpTSTwo  & \LimCombExpTSTwo \\
\hline
\hline
\end{tabular}
\end{center}
\end{table}

\section{Limits on Technicolor} 
\subsection*{LSTC model} 

The Low-scale Technicolor (LSTC) model~\cite{TC2,TC6,Eichten:2012br} 
postulates the existence of vector (\rhot , \omegat ) and axial (\at ) techni-mesons,
in addition to light techni-pions (\pit ).
Due to techni-isospin symmetry,  \rhot\ and  \omegat\ are nearly degenerate in mass. 
Therefore this analysis searches for a combination of  \rhot\ and \omegat , with 
 \omegat\ being the dominant component since its 
 branching fraction to dileptons is approximately one order of magnitude larger than that of the \rhot .
In this work, the 
 LSTC parameters are chosen to be the same as in ref.~\cite{Eichten:2012br} (in particular, the LSTC parameter $\sin\chi = 1/3$)
and the mass of the \at\ state is assumed to be 10\% higher than that of \rhot.

Limits are computed on  $\xbr$ for the decay of the techni-mesons to dilepton final states.  
When building the signal templates, it is assumed that the 
mass splitting is  $M_{\rhot}-M_{\pit}=M_W$. 
Negative interference contributions are neglected.
The intrinsic widths of the \rhot , \omegat\ and \at\ resonances are 
much smaller than the experimental resolution.
The resulting limits on the \rhot/\omegat\ mass are displayed in table~\ref{tab:limits_LSTC}.
\begin{table}[tbp]
\caption{
The observed and expected 95\% CL lower limits on the mass of the \rhot/\omegat\ in the $M_{\rhot}- M_{\pit}=M_W$ hypothesis for the \ee\ and \mumu\ channels separately and for their combination.
}
\label{tab:limits_LSTC}
\begin{center}
\begin{tabular}{l|ccc}
\hline
\hline
                     & $\rhot/\omegat \to \ee$    & $\rhot/\omegat \to \mumu$   & $\rhot/\omegat \to \ll$ \\
\hline			 	 	 	 	  	 
Observed limit [TeV] & \LimitElectronLSTC         & \LimitMuonLSTC              & \LimitCombinedLSTC  \\
Expected limit [TeV] & \LimitElectronExpLSTC      & \LimitMuonExpLSTC           & \LimitCombinedExpLSTC \\
\hline
\hline
\end{tabular}
\end{center}
\end{table}

\begin{figure}[tbp]
 \centering
  \includegraphics[width=0.7\columnwidth]{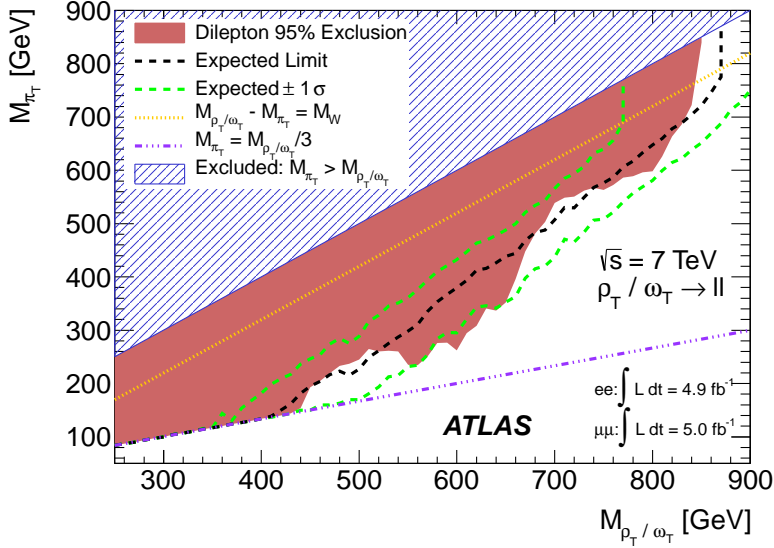}
 \caption{
The 95\% CL excluded region (in red) in the plane \pit\ mass as a function of the \rhot/\omegat\ mass, assuming $M_{\at}=1.1\times M_{\rhot /\omegat}$, for the combination of dielectron and dimuon channels.  
The dotted line corresponds to  $M_{\rhot /\omegat}-M_{\pit}=M_W$. 
The black dashed line shows the expected limit, with  the green dashed lines showing the $\pm 1 \sigma$ bands. 
The blue hashed region in which  $M_{\pit} > M_{\rhot /\omegat}$ is excluded by theory. 
This search is insensitive in the region below the purple dashed-dotted line ($M_{\pit}<M_{\rhot/\omegat}/3$). 
}
\label{fig:LSTC_2D}
\end{figure}

The \xbr\ limits are then translated into exclusion regions in the 
$(M_{\rhot /\omegat} , M_{\pit})$ 
plane, shown in figure~\ref{fig:LSTC_2D}.
The notation $\rhot /\omegat$ indicates the combination of the two resonances. 
The mass splitting between $\rhot$ and $\pit$ determines  
whether decay modes such as $\rhot\to W\pit$ or multi-$\pit$ are allowed kinematically. 
Therefore the choice of the value of the mass of \pit\  has an impact on the ratio between the \at\ and \rhot\ cross sections.
Another foundational assumption of the LSTC model is that the walking TC gauge coupling causes an enhancement of $M_{\pit}$ relative to $M_{\rhot}$ 
and the other vector meson masses.  This tends to close off the $\rhot \to \pit \pit$ decay channel and, even more strongly, 
closes off the \omegat\ and $\at \to 3 \pit$ channels~\cite{Lane:1989ej}.
If $M_{\omegat}>  3 M_{\pit}$, the $\omegat \to  \pit \pit \pit$ channel opens up and quickly becomes the dominant decay mode of \omegat.  
Therefore the dilepton branching fractions become substantially smaller 
and there is no sensitivity in the $M_{\pit}<M_{\rhot/\omegat}/3$ region in the dilepton channel.

\subsection*{MWT model} 

 The Minimal Walking Technicolor (MWT)~\cite{TC3,TC4,TC5} model can be characterized by the following parameters:
 \begin{myitemize}
 \item bare axial and vector masses: $M_A$ and $M_V$;
 \item \gtilde, the strength of the spin-1  resonance interaction;
 \item $M_H$, the Higgs boson mass;
 \item $s$, the coupling of the Higgs boson to composite spin-1 states; 
 \item $S$, the $S$-parameter obtained using the zeroth Weinberg Sum Rule~\cite{Appelquist:1998xf,Belyaev:2008yj}. 
 \end{myitemize}
 
This model predicts only two resonances,  \Rone\ and \Rtwo. $M_{\Rone}$ is lower than $M_{\Rtwo}$ and generally very close to $M_A$. 
In contrast to LSTC, \Rone\ and \Rtwo\  are neither degenerate nor very narrow.
In this work, three free parameters have been set to  
$M_H=200$~GeV,  $s=0$, and $S=0.3$,
 following the recommendation from ref.~\cite{Andersen:2011nk}. 
The mass of the lightest resonance, $M_{\Rone}$, is then scanned in steps of 100~GeV for various values of \gtilde . 
For each choice of \gtilde\ and $M_{\Rone}$, the values of $M_{\Rtwo}$, $M_A$ and $M_V$ are uniquely determined. 

\begin{figure}[tbp]
 \centering
  \includegraphics[width=0.6\columnwidth]{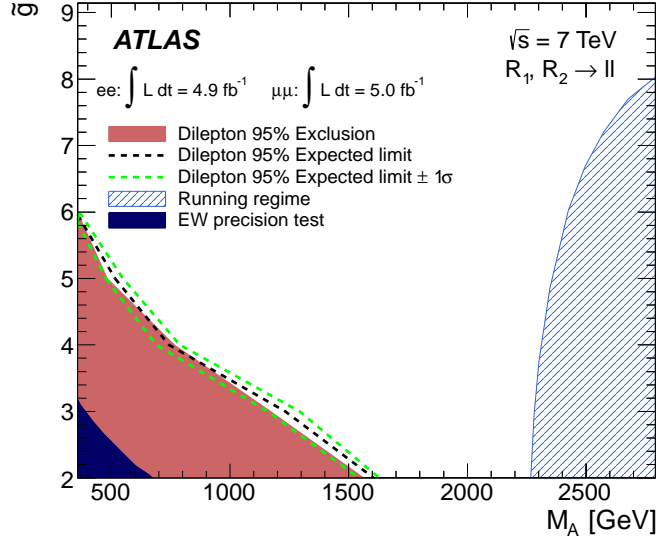}
 \caption{
 Bounds in the ($M_A$, $\tilde{g}$) plane of the MWT parameter space: 
(i) The electroweak precision measurements exclude the dark area in the bottom left corner. 
(ii) The requirement to stay in the walking regime excludes the hatched area in the right corner. 
(iii) The red area (black dashed line) shows the observed (expected) exclusion at 95\% CL in the dilepton channel. The green dashed lines show the $\pm 1 \sigma$ bands of the expected exclusion limit. 
}
\label{fig:MWT_2D}
\end{figure}

Limits on $\xbr (\Ronetwo\to \ll)$ 
are first set as a function of $M_{\Rone }$ assuming $\gtilde=2, 3, 4, 5, 6$, where 
the notation  \Ronetwo\  indicates that both resonances are taken into account in the spectrum.
They are then translated into a 95\% CL exclusion area in the $(M_A,\gtilde)$ plane, 
as shown in figure~\ref{fig:MWT_2D} and table~\ref{tab:combinedLimitsMA}.
The limits from the Tevatron, as well as
the theoretical limits, including the requirement to stay in the walking regime, 
are described in detail in ref.~\cite{Belyaev:2008yj}.  
Note that the edge of the excluded area varies only very weakly as a function of $s$ and $M_H$, 
so a Higgs boson mass of $\approx 125$~GeV would not change the results significantly.
A theoretical re-interpretation of the CMS results from \wp\ boson searches~\cite{Chatrchyan:2011dx}, 
in terms of the parameters $M_A$ and \gtilde, is described in ref.~\cite{Andersen:2011nk}.
\begin{table}[tbp]
\caption{
The observed and expected 95\% CL lower limits on the $M_A$ parameter with varying coupling \gtilde. 
Both lepton channels are combined.
}
\label{tab:combinedLimitsMA}
\begin{center}
\begin{tabular}{l|ccccc}
\hline
\hline
\gtilde              &  6 & 5 & 4 & 3 & 2  \\
\hline			 	 	 	 	  	 
Observed limit [GeV] & \LimCombMASix    & \LimCombMAFive    & \LimCombMAFour    &\LimCombMAThree   &\LimCombMATwo   \\
Expected limit [GeV] & \LimCombExpMASix & \LimCombExpMAFive & \LimCombExpMAFour &\LimCombExpMAThree&\LimCombExpMATwo\\
\hline
\hline
\end{tabular}
\end{center}
\end{table}

%% file: limits_with_interf.tex
\section{Limits on spin-1 Kaluza-Klein $S^1/Z_2$ bosons } 
\label{sec_KK}

The model proposed in ref.~\cite{Rizzo:1999en} assumes a single extra spatial dimension with size of order 1 TeV$^{-1}$, compactified onto an $S^1/Z_2$ orbifold.
In the minimal model considered here, all of the SM fermions are on the same orbifold point. 
The model is completely specified by a single parameter, the compactification scale, which drives the masses of the KK modes.
As for the case of \zpssm , this type of model can be classified as sequential to the Standard Model since 
the KK couplings are kept SM-like, although enhanced by a factor of $\sqrt{2}$.
However, contrary to any of the \zp\ models, the interference with \dy\ is very strong 
and is a potentially distinctive feature of this type of model~\cite{Rizzo:2009pu,Bella:2010sc}.

Because of the strong destructive interference effects mentioned above, it is not possible to put limits on \xbr\ as done for the preceding models.
Instead, a coupling strength $g$ is introduced that multiplies the fermion couplings, 
$g_{\lambda_{f}}^{X}$, where $X$ stands for the new massive \zkk\ resonance, 
and $\lambda_{f}$ can be the helicity coupling, $\lambda_{f}=$L,R, as done in ref.~\cite{Bella:2010sc}.
The resulting differential cross section, after the $g_{\lambda_{f}}^{X} \longrightarrow g\times g_{\lambda_{f}}^{X}$ transformation, is  

\begin{equation*}
	\frac{d\sigma}{ds} \propto \left|{\dy} 
+ \frac{ g_{\lambda_{q}}^{X} g_{\lambda_{\ell}}^{X} }{ s-m_{X}^2 + i\Gamma_{X} m_{X} }\right|^2  \longrightarrow \left|{\dy} 
+ g^2\frac{ g_{\lambda_{q}}^{X} g_{\lambda_{\ell}}^{X} }{ s-m_{X}^2 + ig^2\Gamma_{X} m_{X} }\right|^2.
\end{equation*}
Flat priors of  $g^4$ and  $g^2$ are used in the limit-setting procedure, as opposed to \xbr\ used earlier.
A flat prior in $g^4$ can be assumed  when the pure \zkk\ cross-section term dominates.
If the interference term between \zkk\ and \dy\ dominates, a flat prior in $g^2$ is better motivated. 
Two-dimensional templates are produced in order to scan the $g$ parameter in the region 0--2.2 and the \zkk\ pole masses 
(\mkk)  between 130~GeV and 6~TeV. 

The strong interference with the \dy\ implies a greater sensitivity to shape distortions, 
especially at the high-end of the mass window, and therefore, 
two more systematic uncertainties which are found to be negligible in the non-interfering channels have to be taken into account here.
First, an uncertainty on the muon momentum resolution, which goes up to 20\%--30\% above 2.5~TeV.
Second, an uncertainty on the extrapolation of the \ttbar\ and diboson backgrounds, due to the fit function choice and
the fit range variation; in the dimuon channel, this uncertainty ranges from 2\% to 6\% in the 2--3~TeV mass range, relative to the full background.
These two uncertainties do not affect the dielectron channel due to a better resolution and to the dominance of
the QCD and \wpjet\ background uncertainties over the \ttbar\ and diboson background uncertainties.

\begin{figure}[!tb]
 \centering
  \includegraphics[width=0.49\columnwidth]{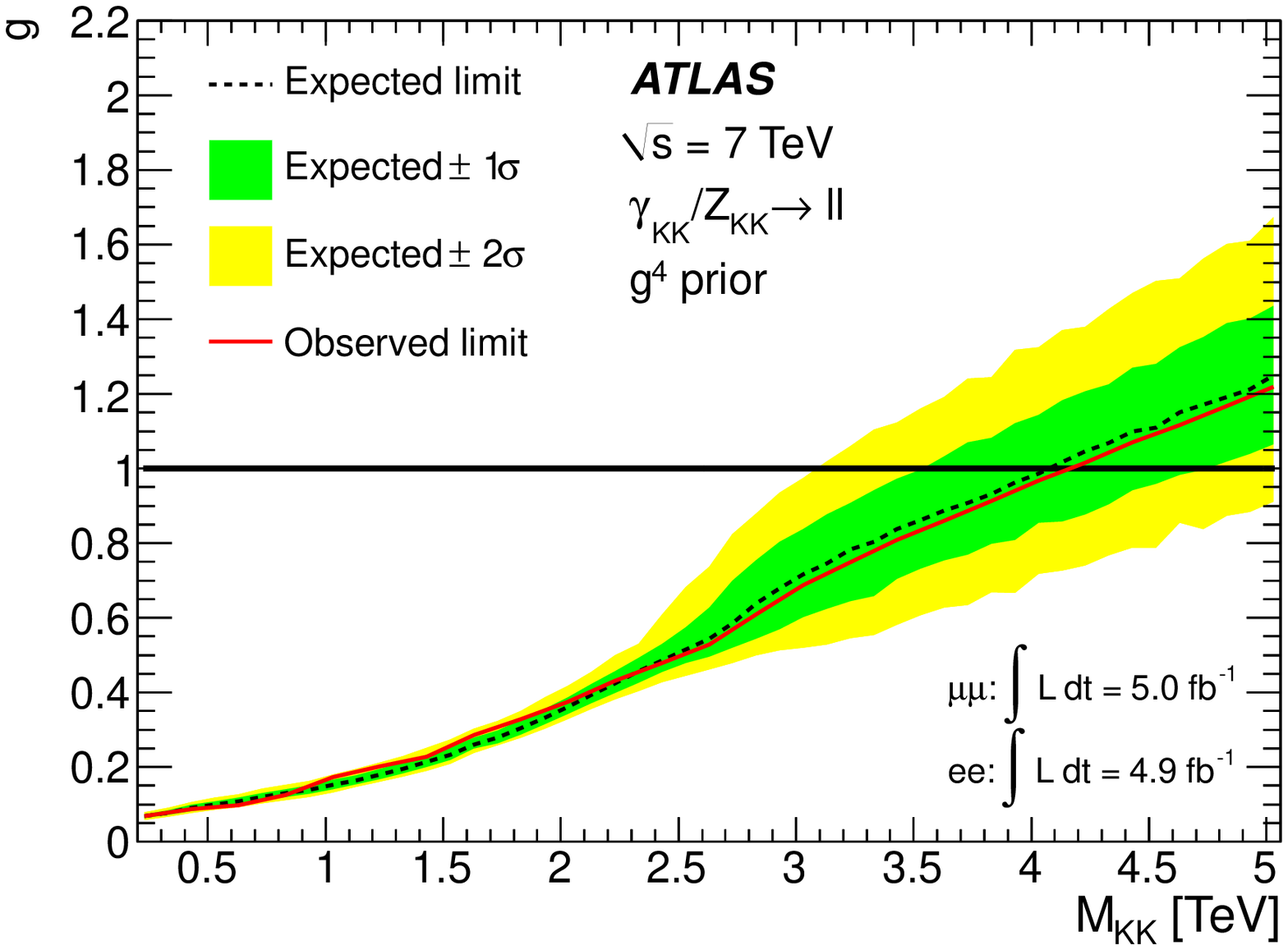}
  \includegraphics[width=0.49\columnwidth]{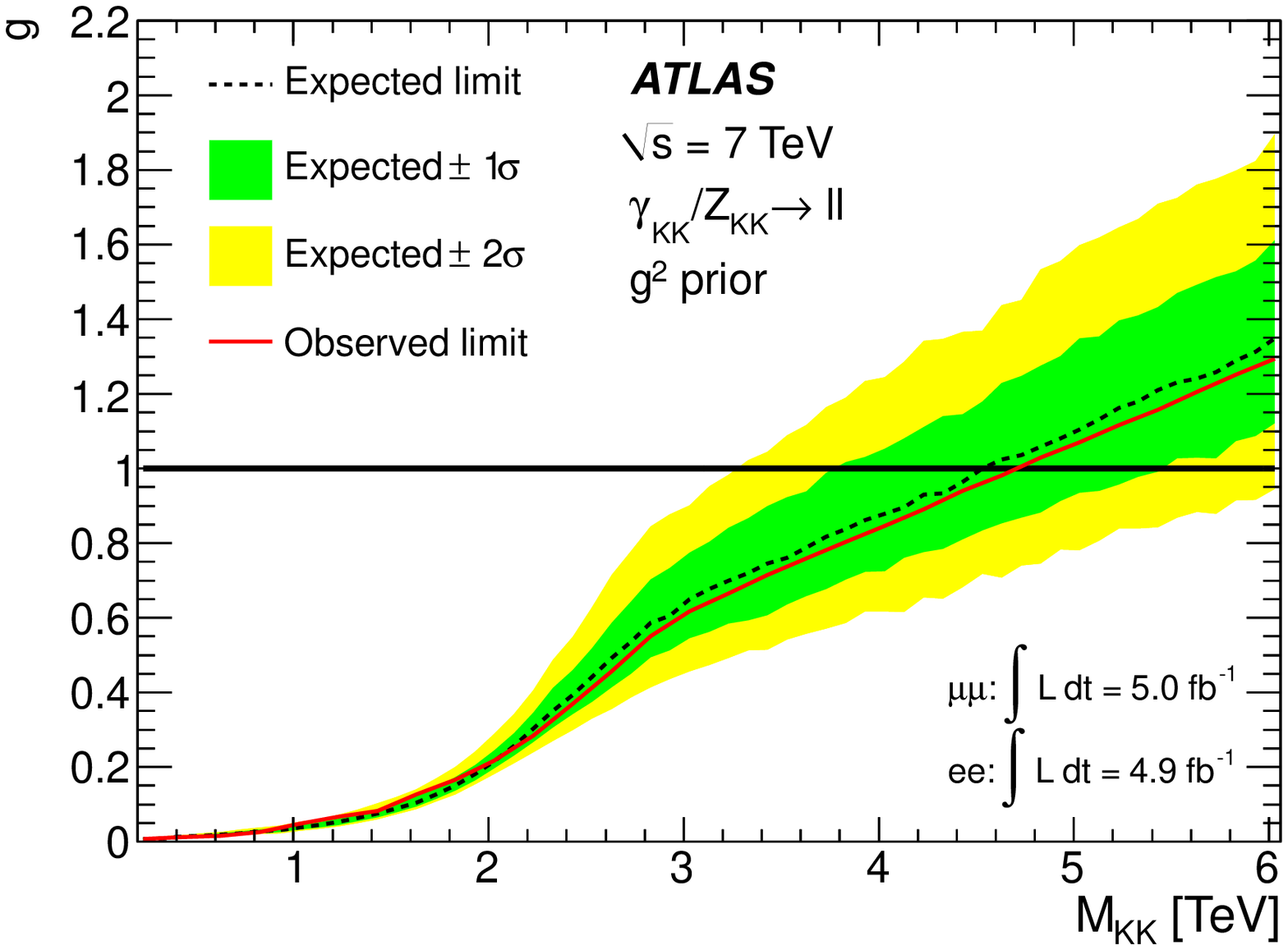}
  \caption{
Expected and observed 95\% CL limits on $g$ as a function of \mkk, for the combination of dielectron and dimuon channels, using a flat prior on $g^4$ (left) and on $g^2$ (right).
}
\label{fig:zkk_interf_lim_comb}
\end{figure}

The observed and expected limits on $g^4$ and $g^2$ are translated into limits on $g$, which are shown as a function of \mkk\ 
in figure~\ref{fig:zkk_interf_lim_comb} for the combination of dielectron and dimuon channels. 
The fast broadening of the expected one and two-sigma bands above 2~TeV is due to the destructive interference becoming the dominant feature of the signal shape.

Lower limits on \mkk\ are derived from the \zkk\ hypothesis ($g = 1$);
they are displayed in table~\ref{tab:limits_ZZ}.
Contrary to the non-interfering case, high-mass candidates in data induce observed limits which are better than expected.
The obtained mass limits are higher than the indirect limits from electroweak precision measurements~\cite{Rizzo:1999en,GG}.

\begin{table}[!hbt]
\caption{
The observed and expected 95\% CL lower limits on the mass of the \zkk\ (i.e. $g=1$) for the \ee\ and \mumu\ channels separately and for their combination. 
}
\label{tab:limits_ZZ}
\begin{center}
\begin{tabular}{l|ccc}
\hline
\hline
                       & $\zkk\to \ee$          & $\zkk\to \mumu$    & $\zkk\to \ll$  \\
\hline
                       &  \multicolumn{3}{c}{$g^4$ prior}       \\
Observed Limit [TeV]   & \LimitElectronKKgf     & \LimitMuonKKgf     & \LimitCombinedKKgf \\
Expected Limit [TeV]   & \LimitElectronExpKKgf  & \LimitMuonExpKKgf  & \LimitCombinedExpKKgf \\
\hline		   
                       &  \multicolumn{3}{c}{$g^2$ prior}       \\
Observed Limit [TeV]   & \LimitElectronKKgsq    & \LimitMuonKKgsq    & \LimitCombinedKKgsq \\
Expected Limit [TeV]   & \LimitElectronExpKKgsq & \LimitMuonExpKKgsq & \LimitCombinedExpKKgsq\\
\hline																
\hline																
\end{tabular}														
\end{center}														
\end{table}

\section{Limits on Minimal \zp\ bosons } 

Limits are also set in the framework of \MM~\cite{Villadoro}.  
In this framework, the coupling of the new boson \zpMM\ to fermions is determined by its coupling to the B--L current, \gbl, 
and its coupling to the weak hypercharge Y, \gy. It is convenient to refer to the ratios $\gbltilde \equiv \gbl / g_Z$ and
 $\gytilde \equiv \gy / g_Z$, where $g_Z$ is the coupling of the SM $Z$ boson defined by $g_Z = 2 M_Z / v$ 
($v = 246$~GeV is the Higgs vacuum expectation value in the SM). 
\gammap~and~$\theta$
are chosen as independent parameters with the following definitions: $\gbltilde=\gammap \cos\theta$, $\gytilde=\gammap \sin\theta$.
The \gammap\ parameter measures the strength of the \zpMM\ boson coupling relative to the SM $Z$ boson coupling,
while $\theta$ determines the mixing between the generators of the B--L and the weak hypercharge $Y$ gauge groups. Specific values of \gammap\ and $\theta$ correspond to \zp\ bosons in various models such as the \zpBL\ boson and \zpthreeR\ boson. 

Signal templates are built which take into account both the interference and the dependence of the \zpMM\ boson width on \gammap\ and $\theta$. 
The coupling to hypothetical right-handed neutrinos and to $W$~boson pairs is neglected.
As for the KK model, the two-dimensional signal templates are made by reweighting the 
simulated \dy\  samples with the ratio of differential cross sections $\delta\sigma(\zpMM + \dy)/\delta\sigma(\dy)$.
For a given value of $\theta$ and for each of the tested pole masses, 
dilepton invariant mass templates are created with varying values of \gammap\ between 0.01 and 2.
The templates at these chosen values of \gammap\ are interpolated to all values of \gammap\  by using a smooth interpolating function in each dilepton
 mass bin.  
The likelihood fit  across all dilepton mass bins finds the most probable value of \gammap\ for each $\theta$ and \zpMM\ boson mass \Mmin. 

\begin{figure}[!tb]
 \centering
  \includegraphics[width=0.7\columnwidth]{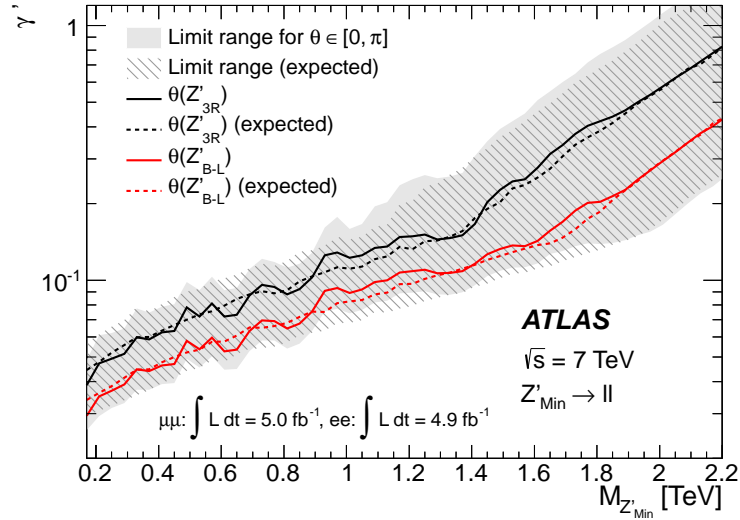}
 \caption{Expected (hatched area and dotted lines) and observed (filled area and solid lines) upper limits on \gammap\ within the \MM\ parameterization. The limits are shown for different test masses and are obtained by 
 combining the dielectron and dimuon channels. The gray band envelops all limit curves, which depend on the choice of $\theta$. The lower boundary corresponds to $\tan{\theta}=  1.43$ and the upper boundary to $\tan{\theta} = -1.19$. 
The limit curves for two representative values of $\theta$ are shown: $\tan{\theta}=0$ and $\tan{\theta}=-2$ which correspond to the \zpBL\ model and the \zpthreeR\ model at specific values of \gammap\ respectively.
}
\label{fig:minimal_limits}
\end{figure}

Systematic uncertainties are applied as in the case of \xbr\ limits. 
Limits are set on the relative coupling strength \gammap\ as a function of the \zpMM\ boson mass, as shown in figure~\ref{fig:minimal_limits}. 
The two $\theta$ values yielding the minimum and maximum cross sections are used to define a band of limits in the (\gammap, \Mmin) plane.
Table~\ref{tab:minimal_limits_mass} shows the range of the lower limits on the \zpMM\ boson mass for representative values of \gammap. The range of the upper limits on \gammap\ for representative values of the \zpMM\ boson mass is shown in table~\ref{tab:minimal_limits_coupling}.

\begin{table}[!htbp]
\caption{Range of the observed and expected 95\% CL lower limits on the \zpMM\ boson mass for $\theta \in [0,\pi]$ and representative values of the relative coupling strength \gammap. Both lepton channels are combined.}
\begin{center}
\begin{tabular}{l|cc} 
\hline
\hline
\gammap & 0.1 & 0.2 \\
\hline
Observed range [TeV] & 0.67-1.43 & 1.11-2.10 \\
Expected range [TeV] & 0.58-1.47 & 1.17-2.07 \\
\hline
\hline
\end{tabular}
\label{tab:minimal_limits_mass}
\end{center}
\end{table}

\begin{table}[!htbp]
\caption{Range of the observed and expected 95\% CL upper limits on the relative coupling strength \gammap\ for $\theta \in [0,\pi]$ and representative values of the \zpMM\ boson mass. Both lepton channels are combined.}
\begin{center}
\begin{tabular}{l|cc} 
\hline
\hline
\zpMM\ mass [TeV] & 1 & 2  \\
\hline
Observed limit & 0.08-0.16 & 0.16-1.10 \\
Expected limit & 0.07-0.15 & 0.17-1.01 \\
\hline
\hline
\end{tabular}
\label{tab:minimal_limits_coupling}
\end{center}
\end{table}

%% file: conclusion.tex

Searches for heavy resonances in the dilepton invariant mass spectrum have been presented.
Proton-proton collisions at a center-of-mass energy of 7~TeV 
with \LumiElectronsFb~\ifb\ in the \ee\  channel
and \LumiMuonsFb~\ifb\ in the \mumu channel have been used.
The observed invariant mass spectra are consistent with the SM expectations. 
Limits are set on the cross section times branching fraction $\xbr$ for spin-1 and spin-2 bosons.
The resulting mass limits are \LimitCombined~TeV for the Sequential Standard Model \zp\ boson, \LimitCombinedPsi$-$\LimitCombinedChi~TeV for various
\esix-motivated \zp\ bosons, and \LimitCombinedG\ (\LimitCombinedGOne)~TeV for a Randall-Sundrum graviton \gstar\
 with the coupling parameter \kovermb\ equal to 0.1 (0.01).
The \gstar\ boson limits are the most stringent to date. 
Experimental limits have also been set on Technicolor models, on \zstar, and for the first time, on Kaluza-Klein modes of electroweak bosons, 
general Minimal Models of \zp\ bosons, and Torsion models in quantum gravity.

%% file: acknowl.tex
We thank Kenneth Lane for useful discussions on details of the LSTC model.

We thank CERN for the very successful operation of the LHC, as well as the
support staff from our institutions without whom ATLAS could not be
operated efficiently.

We acknowledge the support of ANPCyT, Argentina; YerPhI, Armenia; ARC,
Australia; BMWF, Austria; ANAS, Azerbaijan; SSTC, Belarus; CNPq and FAPESP,
Brazil; NSERC, NRC and CFI, Canada; CERN; CONICYT, Chile; CAS, MOST and NSFC,
China; COLCIENCIAS, Colombia; MSMT CR, MPO CR and VSC CR, Czech Republic;
DNRF, DNSRC and Lundbeck Foundation, Denmark; EPLANET and ERC, European Union;
IN2P3-CNRS, CEA-DSM/IRFU, France; GNAS, Georgia; BMBF, DFG, HGF, MPG and AvH
Foundation, Germany; GSRT, Greece; ISF, MINERVA, GIF, DIP and Benoziyo Center,
Israel; INFN, Italy; MEXT and JSPS, Japan; CNRST, Morocco; FOM and NWO,
Netherlands; RCN, Norway; MNiSW, Poland; GRICES and FCT, Portugal; MERYS
(MECTS), Romania; MES of Russia and ROSATOM, Russian Federation; JINR; MSTD,
Serbia; MSSR, Slovakia; ARRS and MVZT, Slovenia; DST/NRF, South Africa;
MICINN, Spain; SRC and Wallenberg Foundation, Sweden; SER, SNSF and Cantons of
Bern and Geneva, Switzerland; NSC, Taiwan; TAEK, Turkey; STFC, the Royal
Society and Leverhulme Trust, United Kingdom; DOE and NSF, United States of
America.

The crucial computing support from all WLCG partners is acknowledged
gratefully, in particular from CERN and the ATLAS Tier-1 facilities at
TRIUMF (Canada), NDGF (Denmark, Norway, Sweden), CC-IN2P3 (France),
KIT/GridKA (Germany), INFN-CNAF (Italy), NL-T1 (Netherlands), PIC (Spain),
ASGC (Taiwan), RAL (UK) and BNL (USA) and in the Tier-2 facilities
worldwide.

%% file: atlas_authlist.tex
\begin{flushleft}
{\Large The ATLAS Collaboration}

\bigskip

G.~Aad$^{\rm 48}$,
T.~Abajyan$^{\rm 21}$,
B.~Abbott$^{\rm 111}$,
J.~Abdallah$^{\rm 12}$,
S.~Abdel~Khalek$^{\rm 115}$,
A.A.~Abdelalim$^{\rm 49}$,
O.~Abdinov$^{\rm 11}$,
R.~Aben$^{\rm 105}$,
B.~Abi$^{\rm 112}$,
M.~Abolins$^{\rm 88}$
% ...
\bigskip

% ...
$^{11}$ Institute of Physics, Azerbaijan Academy of Sciences, Baku, Azerbaijan\\
$^{12}$ Institut de F\'{i}sica d'Altes Energies and Departament de F\'{i}sica de la Universitat Aut\`{o}noma de Barcelona and ICREA, Barcelona, Spain\\
% ...
$^{21}$ Physikalisches Institut, University of Bonn, Bonn, Germany\\
% ...
$^{48}$ Fakult\"{a}t f\"{u}r Mathematik und Physik, Albert-Ludwigs-Universit\"{a}t, Freiburg, Germany\\
$^{49}$ Section de Physique, Universit\'{e} de Gen\`{e}ve, Geneva, Switzerland\\
% ...
$^{88}$ Department of Physics and Astronomy, Michigan State University, East Lansing MI, United States of America\\
% ...
$^{105}$ Nikhef National Institute for Subatomic Physics and University of Amsterdam, Amsterdam, Netherlands\\
% ...
$^{111}$ Homer L. Dodge Department of Physics and Astronomy, University of Oklahoma, Norman OK, United States of America\\
$^{112}$ Department of Physics, Oklahoma State University, Stillwater OK, United States of America\\
% ...
$^{115}$ LAL, Universit\'{e} Paris-Sud and CNRS/IN2P3, Orsay, France\\
% ...
\end{flushleft}
